\documentclass[twocolumn,epjc3]{svjour3}  
\smartqed  
\usepackage{enumerate}   
\RequirePackage{graphicx}
\RequirePackage{latexsym}
\RequirePackage[numbers,sort&compress]{natbib}
\RequirePackage[colorlinks,citecolor=blue,urlcolor=blue,linkcolor=blue]{hyperref}

\usepackage[british]{babel}
\usepackage{amsfonts,amssymb,amsbsy}
\usepackage{amsmath,latexsym,dcolumn}
\usepackage{pstricks}
\usepackage{slashed}
\usepackage{ulem}

\newcommand{\Dslash}{\ensuremath{D\hspace{-1.2ex} /}}
\def\g#1{g_{\mathrm{#1}}}

\journalname{Eur. Phys. J. C}

\begin{document}

\title{Finite size effects in strongly interacting matter at zero chemical potential from Polyakov loop Nambu-Jona-Lasinio model in the light of lattice data
}

\titlerunning{Finite size effects in strongly interacting matter at zero chemical potential}        

\author{ A. G. Grunfeld \thanksref{e1,addr1,addr2}
        \and
        G. Lugones \thanksref{e2,addr3} 
}

\thankstext{e1}{e-mail: grunfeld@tandar.cnea.gov.ar}
\thankstext{e2}{e-mail: german.lugones@ufabc.edu.br}

\authorrunning{}

\institute{CONICET, Godoy Cruz 2290, (C1425FQB) Buenos Aires, Argentina.  \label{addr1}
           \and
           Departamento de F\'\i sica, Comisi\'on Nacional de
 Energ\'{\i}a At\'omica, Av.Libertador 8250, (1429) Buenos Aires, Argentina. \label{addr2}
           \and
           Universidade Federal do ABC, Centro de Ci\^encias Naturais e Humanas, Avenida dos Estados 5001- Bang\'u, CEP 09210-580, Santo Andr\'e, SP, Brazil.   \label{addr3}
}

\date{Received: date / Accepted: date}

\maketitle

\begin{abstract}
We study finite volume effects within the Polyakov loop Nambu-Jona-Lasinio model for two light and one heavy quarks at vanishing baryon chemical potential and finite temperatures. 
We include three different Polyakov loop potentials and ensure that the predictions of our effective model in bulk are compatible with lattice QCD results. 
Finite size effects are taken into account by means of the Multiple Reflection Expansion formalism. 
We analyze several thermodynamic quantities including the interaction measure, the speed of sound, the surface tension, and the curvature energy and find that they are sensitive to finite volume effects, specially for systems with radii below $\sim 10$ fm and temperatures around the crossover one. 
For all sizes, the system undergoes a smooth crossover.  The chiral critical temperature decreases by around $5 \%$ and the deconfinement temperature  by less than a $2 \%$ when the radius goes from infinity to 3 fm. Thus, as the drop's size decreases, both temperatures become closer. 
The surface tension is dominated by the contribution of strange quarks  and the curvature energy  by $u$ and $d$ quarks. At large temperatures both quantities grow proportionally to $T^{3/2}$ but saturate to a constant value at low $T$. 

\end{abstract}

\section{Introduction}

Understanding the hadron-quark phase transition is still a challenge from both the theoretical and experimental points of view. The framework for describing it is provided by Quantum Chromodynamics (QCD), which is the fundamental theory of strong interactions.  However, the nonperturbative character of QCD at low  energies makes extremely difficult to solve it in the regime of intermediate temperatures and chemical potentials, although lattice methods had a huge progress in the last years \cite{lattice1,lattice2,lattice3,lattice4,lattice5}. In this context, effective models such as the Nambu-Jona-Lasinio (NJL) model \cite{njl1,njl2,njl3,buballa} are very useful because they can address many aspects of the QCD phase diagram without computational shortcomings at finite chemical potentials. The NJL model has many similarities with the full QCD theory but does not take into account the property of confinement, since quarks interact each other via pointlike interactions without exchanged gluons.
Thus, in order to obtain a more realistic description, taking into account the quark confinement at low energies, the Polyakov loop was introduced in the NJL model  ~\cite{Meisinger1996}, leading to the so called Polyakov loop NJL (PNJL) model (see also \cite{Fukushima2004}).
From this widely studied effective QCD model, many properties of strongly interacting matter can be obtained, such as its phase diagram~\cite{fukupd1,fukupd2,fukupd3}.  

On the other hand, a comprehension of finite size properties is very important for situations where the deconfinement transition occurs over a finite volume as in relativistic heavy ion collisions and neutron stars. The strongly interacting matter formed in a heavy-ion collision is finite in volume, and its size depends on the size of the colliding nuclei, the collision center of mass energy, and the centrality of the collision. In neutron stars, a deconfinement transition to quark matter is possible and a hybrid star or a strange quark star can be formed.  The conversion of the star is expected to start with the nucleation of small quark matter drops  \cite{Lugones2010,Bombaci2009,Mintz2010,Lugones:2015bya} which subsequently grow at the expenses of the gravitational energy extracted from the contraction of the object and/or through a strongly exothermic combustion process. Quark matter droplets with a variety of geometrical forms can also arise within the mixed hadron-quark phase that is expected to form inside hybrid stars if global charge neutrality is allowed \cite{Glendenning}. Also, the most external layers of a strange star  may fragment into a charge-separated mixture, involving positively-charged strange droplets (strangelets) immersed in a negatively charged sea of electrons, forming a crystalline solid crust \cite{Jaikumar2006}.

In the past years, many theoretical studies of finite-volume effects have been performed based on the NJL model ~\cite{Abreu2006, Yasui2006, Abreu2011}. However, studies within the PNJL model are more recent \cite{Cristoforetti2010,Bhattacharyya2013, Bhattacharyya2015}.
To incorporate finite-size effects different procedures have been employed,  such as Monte Carlo simulations \cite{Cristoforetti2010}, a renormalization group approach \cite{Tripolt2014}, and the implementation of a low momentum cutoff $\Lambda$ on the integration of the thermodynamic potential
density of the PNJL model \cite{Pan2017}. 

In the present work we use a different approach for the inclusion of finite size effects, known as  Multiple Reflection Expansion (MRE) formalism  \cite{Balian1970}. 
First,  different thermodynamic quantities calculated within our effective model in the bulk (including three different Polyakov loop potentials) are compared to the corresponding lattice QCD results. This is necessary as a starting point to check the validity of our model. Then, we study the relevance of finite size effects on many properties of strongly interacting matter and analyze how they deviate from the bulk case. 

A comparison with lattice QCD is always important to calibrate effective models, 
that can be later extrapolated to a higher density regime.
For example, the effective model can be used to explore finite-volume effects in a regime where they are known to be essential, such as in relativistic heavy ion collisions. 
Additionally, some results could be of interest for the analysis of the cosmological quark-hadron transition, which occurred in the early Universe about 10 $\mu$s after the Big Bang, when a hot unconfined quark-gluon plasma was converted, as the Universe expanded and cooled, into a confined hadronic phase.

The paper is organized as follows. In Sec. \ref{sec:PNJL_bulk} we review the PNJL model in bulk for different Polyakov loop potentials and in Sec. \ref{sec:PNJL_MRE} we introduce finite size effects through the MRE formalism. Our results are presented in Sec. \ref{sec:results} where we analyze the behavior of several thermodynamic quantities such as the chiral critical temperature, the deconfinement temperature, the constituent masses, the interaction measure, the pressure, the energy density, the entropy density, the speed of sound, the surface tension, and the curvature energy for different system sizes. Finally, we present our conclusions in Sec. \ref{sec:conclusions}.

\section{The PNJL model in the bulk}
\label{sec:PNJL_bulk}

The Lagrangian of the Polyakov loop extended  $SU(3)_f$ NJL model including the six-quark 't Hooft interaction reads
\begin{eqnarray}
  \label{eq:pnjl}
  {\cal L} &=& \bar q \left( i \Dslash - \hat m \right) q +\frac{
    \g{S}}{2}\sum_{a=0}^{N_f^2-1} \left[ (\bar q \lambda^a q)^2 +(\bar q
    i\gamma_5\lambda^a q)^2\right]\nonumber \\
  && +\  \g{D} \left[ \det \left(\bar q (1-\gamma_5) q \right) +\det
    \left(\bar q (1+\gamma_5)q \right)\right] \nonumber \\
  &&  -\ {\cal U} (l ,\bar l;T) \ ,
\end{eqnarray}
where $q = (u,d,s)$ represents the three flavor quark field with three colors and $\hat m = \mathrm{diag}(m_u, m_d, m_s)$ stands for the current quark mass matrix. 
We assume the $SU(2)_V$ isospin symmetry limit in which $ m_u = m_d $. 
The covariant derivative in the fermion kinetic term couples a temporal background gauge field, the Polyakov loop, to the quark fields {through $D_\mu = \partial_\mu -i A_\mu$ with $A_\mu = \delta_\mu^0 A_0$ in Polyakov gauge, and $A_0 = -i A_4$.} 
Here, we used the notation $A_\mu = g A^a_\mu \lambda^a/2$ with $g$ the $SU(3)_c$ gauge coupling. The $\lambda^a´s$ stand for the Gell-Mann
matrices with $\lambda^0 = \sqrt{2/3}\ {\bf 1}$ in flavor space. The four-quark interaction coupling in the (pseudo)scalar channel is
denoted by $\g{S}$ and the six-quark 't Hooft interaction coupling, induced by instantons, is labeled by $\g{D}$. The latter one breaks the axial $U_A(1)$ symmetry. Finally, the above Lagrangian includes an effective potential ${\cal U} ( l ,\bar l;T)$
that accounts for gauge field self-interactions and is a function of the temperature $T$ and the normalized color-traced Polyakov loop expectation value and its Hermitean conjugate, defined by
\begin{equation}
  l = \langle \mathrm{tr_c} L \rangle /N_c, \qquad  
  \bar l = \langle \mathrm{tr_c} L^\dagger \rangle /N_c, 
\end{equation}
where the Polyakov loop $L$ is an $N_c \times N_c$ matrix in color space, as a function of $A_4$. The explicit form of the Polyakov loop potential
${\cal U} (l ,\bar l;T)$ will be discussed in Sec. \ref{subsec:Polyakov_potentials}.

\subsection{The thermodynamic potential}

Different thermodynamic properties of our model can be obtained from the thermodynamic potential in the mean-field approximation (MFA).
The thermodynamic grand potential $\Omega(T,\mu)$ of the PNJL model in the
MFA has been largely considered in the
literature, see e.g.~\cite{Ratti:2005jh, Fukushima:2008wg}. Based on \cite{Fukushima:2008wg} we write the thermodynamic grand potential per unit volume as follows
\begin{equation}
\Omega_{PNJL} = \Omega_{cond} + \Omega_{zero} + \Omega_{quark} - \Omega_{vac} +
{\cal U} (l ,\bar l;T).
\label{eq:omegafull}
\end{equation}
The first term is the condensation energy, that contains the contribution of the scalar four-quark interaction proportional to $\g{S}$ plus the six-quark 't Hooft interaction, proportional to $\g{D}$. In  the MFA this term depends on the three condensates $\langle{\bar{u}}u\rangle$, $\langle{\bar{d}}d\rangle$ and $\langle{\bar{s}}s\rangle$ as follows
\begin{equation}
\Omega_{cond} = g_S \left[ \langle{\bar{u}}u\rangle^2 + \langle{\bar{d}}d\rangle^2 +
\langle{\bar{s}}s\rangle^2 \right] + 4 g_D
\langle{\bar{u}}u\rangle \langle{\bar{d}}d\rangle \langle{\bar{s}}s\rangle. 
\end{equation}
The zero point energy
\begin{equation}
\Omega_{zero} = -2 N_c \int^{\Lambda} \sum_i \frac{d^3p}{(2 \pi)^3} \epsilon_i(p)
\end{equation}
is clearly divergent. Since the PNJL model is non-re\-nor\-ma\-lizable, the
zero-point energy contribution requires an ultraviolet cutoff $\Lambda$. 
The quark quasiparticle energies are denoted by $\epsilon_i(p) = \sqrt{p^2 + M_i^2}$ where the constituent quark masses $M_i$ for flavors $i=u, d, s$ are:
\begin{eqnarray}
M_u &=& m_u - 2 g_S \langle{\bar{u}}u\rangle - 4 g_D \langle{\bar{d}}d \rangle \langle {\bar{s}}s \rangle, \\
M_d &=& m_d - 2 g_S \langle{\bar{d}}d\rangle - 4 g_D \langle {\bar{s}}s \rangle \langle {\bar{u}}u \rangle,  \\
M_s &=& m_s - 2 g_S \langle{\bar{s}}s\rangle - 4 g_D \langle {\bar{u}} u \rangle \langle {\bar{d}} d \rangle ,
\end{eqnarray}
being $m_i$ the current quark masses. 

The  $\Omega_{quark}$ term is ultraviolet finite and hence no momentum cutoff is imposed on it. It contains the coupling between the chiral condensates and the Polyakov loop $L$, and reads \cite{Fukushima:2008wg}: 
\begin{eqnarray}
\Omega_{quark} &=& -2 T \sum_i \int \frac{d^3p}{(2 \pi)^3}  \ln  
\det [ 1 + L e^{-\frac{\epsilon_i - \mu_i}{T}} ] \nonumber \\
&& -2 T \sum_i \int \frac{d^3p}{(2 \pi)^3} \ln  \det [1 +  L^\dagger e^{-\frac{\epsilon_i + \mu_i}{T}} ].
\label{omega:quark}
\end{eqnarray}
As shown in \cite{Fukushima:2008wg}, taking an average of the $3 \times 3$ determinant we obtain:
\begin{equation}
\label{eq:Oql}
\Omega_{quark} = -2 T \sum_i \int \frac{d^3p}{(2 \pi)^3}  [\ln \langle \det f_i^- \rangle +  \ln  \langle \det f_i^+ \rangle],
\end{equation}
where 
\begin{eqnarray}
 \langle \det f_i^- \rangle & = & 1 + e^{-3(\epsilon_i - \mu_i)/T} + 3 \, l  e^{-(\epsilon_i - \mu_i)/T} \nonumber \\
&& + 3 \bar {l}  e^{-2(\epsilon_i - \mu_i)/T} ,  \label{eq:fmen}  \\
\langle \det f_i^+ \rangle & = &1 + e^{-3(\epsilon_i + \mu_i)/T} + 3 \bar l  e^{-(\epsilon_i + \mu_i)/T} \nonumber \\
&& + 3 l  e^{-2(\epsilon_i(p) + \mu_i)/T} .   \label{eq:fmas}
\end{eqnarray}

The fourth contribution in Eq. (\ref{eq:omegafull}) is a constant  $\Omega_{vac} \equiv - P_{vac}$, which is {usually} introduced in order to obtain a vanishing pressure at vanishing temperature and chemical potential. {We will discuss the procedure for fixing $P_{vac}$ and its effect on the thermodynamic quantities in the next section.} 

Finally, the term ${\cal U} (l ,\bar l;T)$ in Eq. (\ref{eq:omegafull}),  represents the pure gluonic effective potential in terms of the Polyakov loop variables, which will be presented below in detail. Notice that the ${\cal U} (l ,\bar l;T)$ potential and $\Omega_{quark}$ are invariant under the simultaneous exchange of $l \leftrightarrow \bar l$ together with $-\mu_i \leftrightarrow + \mu_i$.
Let us remark that for three quark flavors the thermodynamic grand potential $\Omega(T,\mu_i)$ generally depends on three independent quark chemical potentials $\mu_i$. As a consequence of the isospin symmetry, the light quark chemical potentials are also degenerated.  In the present work, we consider quark matter to be symmetric and define a common chemical potential $\mu \equiv \mu_u = \mu_d = \mu_s$. Moreover, since we want to compare our results in the bulk with lattice QCD results we will work at finite temperature and vanishing chemical potential.

In order to obtain the dependence of the order parameters on the temperature and the chemical potential, one has to solve the following set of coupled equations: 
\begin{eqnarray}
\frac{\partial \Omega_{PNJL}}{\partial \langle{\bar{u}}u \rangle} &=& 
\frac{\partial \Omega_{PNJL}}{\partial \langle{\bar{d}}d \rangle} = 
\frac{\partial \Omega_{PNJL}}{\partial \langle{\bar{s}}s \rangle} = 0, \\
\frac{\partial \Omega_{PNJL}}{\partial l} &=& 
\frac{\partial \Omega_{PNJL}}{\partial {\bar{l}}} = 0.
\end{eqnarray}

These conditions are consequences from the fact that the thermodynamically consistent solutions
correspond to the stationary points of $\Omega_{PNJL}$ with respect to $\langle{\bar{u}}u\rangle$, $\langle{\bar{d}}d\rangle$, $\langle{\bar{s}}s\rangle$, $l$ and ${\bar{l}} $.

\subsection{Polyakov loop potentials}
\label{subsec:Polyakov_potentials}

The choice of the effective Polyakov loop potential ${\cal U}$ is not unique. 
In general, it can be constructed from the center symmetry of the pure-gauge sector. The required parameters can be extracted from pure gauge lattice data at $\mu=0$ \cite{Pisarski:2000eq}. 
Among several possible choices, see e.g.~\cite{BJ}, we will use the following effective Polyakov loop potentials:

\begin{enumerate}[(i)]

\item \textit{Logarithmic potential}: the logarithmic ansatz presented in~\cite{BJ} is:
\begin{eqnarray}
\frac{{\cal U}_L}{T^4} &=& -\frac{a(T)}{2}  l \bar{l} +  b(T) \ln[1 - 6 l \bar{l}  - 3 (l \bar{l})^2 \nonumber \\
&& + 4 (l^3  + \bar{l}^3)], 
\end{eqnarray}
where $a(T)$ and $b(T)$ are defined by \cite{Roessner:2006xn}:
\begin{eqnarray}
a(T) & = & a_0 + a_1 (T_0/T) + a_2 (T_0/T)^2, \\
b(T) & = & b_3 (T_0/T)^3 ,
\end{eqnarray}
with $a_0 = 3.51$,  $a_1= -2.47$, $a_2 = 15.2$ and $b_3 = -1.75$.

\item \textit{Polynomial potential}: Another choice is \cite{Ratti:2005jh}:
\begin{equation}
\frac{{\cal U}_P}{T^4} =   -\frac{b_2(T)}{2}\, l \bar{l}  -\frac{b_3}{6}\,(l^3 + \bar{l}^3)\, + \frac{b_4}{4}\,(l \bar{l})^2 ,
\end{equation}
where 
\begin{eqnarray}
b_2(T) & = & a_0 + a_1 (T_0/T) + a_2 (T_0/T)^2 \nonumber \\
&& + a_3 (T_0/T)^3, 
\end{eqnarray}
with $a_0 = 6.76$,  $a_1= -1.95$,  $a_2 = 2.625$, $a_3 = -7.44$,  $b_3 = 0.75$ and  $b_4 = 7.5$.
In the absence of dynamical quarks, in a pure gauge sector, 
one expects a deconfinement temperature $T_0 = 270$ MeV.
Nevertheless, in \cite{Schaefer:2007pw} it has been shown that in the presence of two light dynamical quarks and a massive strange one, this temperature is rescaled to about 187 MeV, with an uncertainty of about 30 MeV. In fact, for $N_f = 2+1$, $T_0 = 187$ MeV and $T_0 = 190$ MeV have been used in \cite{BJ} and in \cite{Bhattacharyya2013} respectively.
Here we use $T_0 = 185$ MeV.

\item \textit{Fukushima potential}: Finally, we will use the strong-coupling inspired version of the effective Polyakov potential with only two parameters $a$ and $b$ proposed by Fukushima \cite{Fukushima:2008wg}
\begin{eqnarray}
\frac{{\cal U}_F}{T^4} & = & -\frac{b}{T^3} [ \, 54 e^{-a/T} l \bar{l} \nonumber \\
 &&  + \ln \{ 1 - 6 l \bar{l} - 3 (l \bar{l})^2 +4( l^3+ \bar{l}^3) \} \, ].
\end{eqnarray}
{The first term (proportional to $l \bar{l} $) reminds the nearest neighbor interaction in the effective action at strong coupling and its  temperature-dependent coefficient controls the deconfinement phase transition temperature.
The logarithmic term comes from the Haar measure of the group integration with respect to the SU(3) Polyakov loop matrix. }
{The parameters $a$ and $b$ are independent of the temperature, the chemical potential and the number of quark flavors $N_f$. 
The parameter $a$ controls only the deconfinement transition temperature and can be determined by the condition that the first-order phase transition in pure gluodynamics takes place at T = 270 MeV, which results in  $a = 664$ MeV. 
On the other hand, the parameter $b$ can be used to control the relative value of the deconfinement and chiral restoration crossover temperatures. Since there is no established prescription for fixing $b$, we shall adopt here two different values. }
First, we consider $b = (196.2 \, \mathrm{MeV})^3$ as suggested in \cite{Fukushima:2008wg,BJ} leading to an almost simultaneous crossover for deconfinement and chiral restoration at a  temperature of $T \simeq 200$ MeV (we call this case ${\cal U}_{F1}$). 
The second choice is $b = (115 \, \mathrm{MeV})^3$ (we call this case ${\cal U}_{F2}$) which gives lower critical temperatures as we will see below.

\end{enumerate}

\subsection{Parametrization}

In order to fully specify the non-local model under consideration we fix the model parameters following Ref. \cite{Noguera:2008cm}.
For comparison with some recent results \cite{Bratovic:2012qs}, we have considered the parameters in \cite{buballa}, $m_u = m_d = 5.5$  MeV, $m_s = 135.7$  MeV, $\Lambda = 631.4$  MeV,  $g_S\cdot\Lambda^2 = 3.67$ and $g_D\cdot\Lambda^5 = -9.29$.

\section{Finite size effects within the MRE formalism}
\label{sec:PNJL_MRE}
Now we are ready to introduce the effects of finite size in the thermodynamic potential. For doing so we consider the MRE formalism (see Refs. \cite{Balian1970,Madsen-drop,Kiriyama1,Kiriyama2} and references therein) which takes into account the modification in the density of states resulting when the system is restricted to a finite domain.   
For the case of a finite spherical droplet the density of states reads: 
\begin{equation}
\rho_{i,\mathrm{MRE}}(p,m_i,R) = 1 + \frac{6\pi^2}{p R} f_{i,S} + \frac{12\pi^2}{(p R)^2} f_{i,C}
\end{equation}
where the surface contribution to the density of states is
\begin{equation}
f_{i,S}  = - \frac{1}{8 \pi} \left(1 -\frac{2}{\pi} \arctan \frac{p}{m_i} \right), 
\end{equation}
and the curvature contribution is given by Madsen's ansatz \cite{Madsen-drop}
\begin{equation}
f_{i,C}  =  \frac{1}{12 \pi^2} \left[1 -\frac{3p}{2m_i} \left(\frac{\pi}{2} - \arctan \frac{p}{m_i} \right)\right] ,
\end{equation}
which takes into account the finite quark mass contribution.

The MRE density of states for massive quarks is reduced compared with the bulk one, and for a range of small momenta becomes negative. This non-physical negative values are removed
by introducing an infrared (IR) cutoff in momentum space \cite{Kiriyama2}. Thus,  we have to perform the following replacement in order to obtain the thermodynamic quantities
\begin{equation}
\int_0^{\Lambda, \infty} \cdots \frac{d^3p}{(2 \pi)^3} \longrightarrow
\int_{\Lambda_{i,\mathrm{IR}}}^{\Lambda,\infty} \cdots \rho_{i,\mathrm{MRE}} \, \frac{d^3p}{(2 \pi)^3}.
\label{MRE}
\end{equation}
The upper integration limit is either infinity or given by a cutoff $\Lambda$. The IR cut-off $\Lambda_{i,\mathrm{IR}}$ is the largest solution of the equation $\rho_{i,\mathrm{MRE}}(p,m_i,R)= 0$ with respect to the momentum $p$. 

After the above replacement, the \textit{full} thermodynamic potential $\mathbf{\Omega}_{\mathrm{MRE}}$ for a finite size spherical droplet reads: 
%

\begin{eqnarray}
\frac{\mathbf{\Omega}_{\mathrm{MRE}}}{V} &=& \Omega_{cond} + {\cal U} (l ,\bar l;T) \nonumber \\
&& - 2 N_c \sum_i \int_{\Lambda_{i, \mathrm{IR}}}^{\Lambda}  
\epsilon_i(p) \, \rho_{i, \mathrm{MRE}}  \frac{d^3p}{(2 \pi)^3} \nonumber \\
&& - 2 T \sum_i \int_{{{\Lambda_{i, \mathrm{IR}}}}}^{\infty}   
\left[\ln \langle \det f_i^- \rangle + \ln \langle \det f_i^+ \rangle \right] \times \nonumber \\
&& \qquad \qquad \times \rho_{i, \mathrm{MRE}}  \frac{d^3p}{(2 \pi)^3}  +   P_{vac} .
\label{fullomega}
\end{eqnarray}
Multiplying on both sides of the last equation by the volume of the quark matter drop, replacing
the area $S= 4\pi R^2$  and the curvature $C=8\pi R$ for a spherical drop,
and rearranging terms we arrive to the following form for $\mathbf{\Omega}_{\mathrm{MRE}}$ 
\begin{equation}
\mathbf{\Omega}_{\mathrm{MRE}} = - P V + \alpha S + \gamma C ,
\label{eq17}
\end{equation}
where the pressure $P$, the surface tension $\alpha$ and the curvature energy density $\gamma$, are defined as in Ref.  \cite{Lugones2011}:
\begin{eqnarray}
P   & \equiv &  - \frac{\partial \mathbf{\Omega}_{\mathrm{MRE}}}{ \partial V } \bigg|_{T, \mu, S, C}  \\ 
&=&  - \Omega_{cond} - {\cal U} (l ,\bar l;T)  + 2 N_c \sum_i \int_{\Lambda_{i, \mathrm{IR}}}^{\Lambda}  
\epsilon_i(p) \frac{dp^3}{(2 \pi)^3} \nonumber \\
&& + 2 T \sum_i \int_{\Lambda_{i, \mathrm{IR}}}^{\infty} 
[\ln \langle \det f_i^- \rangle + \ln \langle \det f_i^+ \rangle ] \frac{dp^3}{(2 \pi)^3} \nonumber \\
&& -   P_{vac},  \nonumber
\label{P} \\
\alpha & \equiv &  \frac{\partial \mathbf{\Omega}_{\mathrm{MRE}}}{ \partial S }
\bigg|_{T, \mu, V, C}   \\
&= & - 2 N_c \sum_i \int_{\Lambda_{i, \mathrm{IR}}}^{\Lambda}   \epsilon_i(p) f_{i, S} \, p dp  \nonumber \\
&& -2 T \sum_i \int_{{{\Lambda_{i, \mathrm{IR}}}}}^{\infty}  
  [\ln \langle \det f_i^- \rangle + \ln \langle \det f_i^+ \rangle ] f_{i, S}\, p dp ,
\label{surfacetension} \nonumber   \\
\gamma &\equiv & \frac{\partial \mathbf{\Omega}_{\mathrm{MRE}}}{ \partial C } \bigg|_{T, \mu, V, S}  \\ 
&=&  - 2 N_c \sum_i \int_{\Lambda_{i, \mathrm{IR}}}^{\Lambda}  \epsilon_i(p) f_{i, C} \, dp  \nonumber \\
&& -2 T \sum_i \int_{\Lambda_{i, \mathrm{IR}}}^{\infty} 
[\ln \langle \det f_i^- \rangle + \ln \langle \det f_i^+ \rangle ] f_{i, C} \, dp .
\nonumber
\label{curvatureenergy}
\end{eqnarray}

As we previously mentioned, the value of $\Lambda_{\mathrm{IR}}$ is the largest root when solving $\rho_{i,\mathrm{MRE}}(p,m_i,R)= 0$ with respect to the momentum $p$, i.e. 
$\Lambda_{\mathrm{IR}}$ changes with $m_i$ and with the drop's radius $R$.

{Finally, we will address some aspects of the present model that deserve a more detailed discussion:} 

\begin{enumerate}[(i)]
\item  In the present treatment finite-size effects enter the fermion loop integral only; i.e. these effects are not considered in the pure Yang-Mills sector.  As a consequence, the Polyakov loop potential remains unchanged and feels volume effects only implicitly through the saddle point equations. {A more detailed analysis is left for future work.}

\item  The conventional procedure for fixing  $P_{vac}$  is to impose that the grand thermodynamic potential $\Omega$ must vanish at zero temperature and vanishing chemical potential for matter in bulk. For the above quoted parametrization, this assumption leads to the value $P_{vac} = 5080 \, \mathrm{MeV  \, fm}^{-3}$. 
Nevertheless, it has been emphasized in previous works \cite{Schertler1999,Pagliara2008,Lenzi:2012xz} that this prescription is no more than an arbitrary way to uniquely determine the EOS of the NJL model without any further assumptions. 
A change in the value of  $P_{vac}$ has no influence on the fittings of the vacuum values for the meson masses and decay constants and thus the standard prescription for $P_{vac}$  is not related to experimental values. 
In fact, different prescriptions for determining $P_{vac}$ have been adopted \cite{Pagliara2008}, including the alternative of taking it as a free parameter \cite{Lenzi:2012xz} as it is usually done within the MIT bag model for the bag constant. 
When studying finite size systems, the standard choice for $P_{vac}$  has an additional issue.  If  $\Omega$  vanishes at $T=\mu=0$ for matter in bulk it will not do so for a finite size, due to the contribution of surface and curvature effects (as can be seen from Eq. \eqref{eq17}). 

As in previous works \cite{Lugones:2013bo,Lugones:2017ft}, we will fix  $P_{vac}$ in the standard way, i.e. setting $\Omega=0$  at $T=\mu=0$ for matter in bulk, and will use this value for any system's size. Nonetheless, it must be emphasized that most of the thermodynamic quantities of relevance here (such as the critical temperatures, the entropy density, the sound speed, the specific heat, the surface tension and the curvature energy) are independent of the choice $P_{vac}$ since they are related to derivatives of  the grand thermodynamic potential $\Omega$. 
The influence of the $P_{vac}$ choice on other thermodynamic quantities will be discussed below.

\end{enumerate}

\section{Results}
\label{sec:results}
In this section we present our numerical results for some thermodynamic properties of bulk and finite size quark matter systems. We will show the dependence of our results on the size of the system as well as for different choices of the Polyakov loop potential.
We work at zero chemical potential to compare our numerical results for the bulk with those from lattice QCD for (2+1)-flavors using the highly improved staggered quark action extrapolated to the continuum limit \cite{Bazavov:2014} (see also \cite{Borsanyi:2014}).
Then we describe our predictions for finite size systems.

\subsection{Chiral  and deconfinement transitions}

\begin{table}[t]
\centering
\caption{ Using the polynomial Polyakov loop potential and taking different values for the drop's radius $R$, we show the chiral critical temperature $T_{\chi}$ of the $u$ and $d$ condensates, the critical deconfinement temperature $T{_d}$ of the Polyakov loop expectation value,  and the temperature $T_*$ below which the drop's pressure becomes negative. {$T_{\chi}$ and $T{_d}$ are independent of the choice of the vacuum pressure $P_{vac}$. $T_*$ is calculated for the standard value $P_{vac} = 5080 \, \mathrm{MeV  \, fm}^{-3}$. } }

\label{table:Polynomial} 
\begin{tabular}{l|cccc}
\hline
   &   & $R$ [fm] &\\
     &   3   &  5    &  10   &  $ \infty$  \\ 
\hline
$T_{\chi}$ [MeV]  \quad    &  177   & 182   &  184  & 186 \\
$T_{d}$ [MeV]     \quad    &  160   & 161   &  162  & 162 \\
$T_*$ [MeV]       \quad    &  155   & 141   &  124  &   0 \\
\hline
\end{tabular}
\centering
\caption{Same as in Table \ref{table:Polynomial} but for the logarithmic Polyakov loop potential.}
\label{table:Logarithmic} 
\begin{tabular}{l|cccc}
\hline
 & & &  $R$ [fm]   &\\
                       &   3   &  5    &  10   &  $ \infty$  \\ 
\hline
$T_{\chi}$ [MeV] \quad      &  181   & 187   &  190  & 192 \\
$T_{d}$ [MeV]    \quad      &  150   & 151   &  152  & 152 \\
$T_*$ [MeV]      \quad      &  157   & 149   &  132  &   0 \\
\hline
\end{tabular}

\centering
\caption{Same as in Table \ref{table:Polynomial} but for the Polyakov loop  potential of Fukushima, version ${\cal U}_{F1}$.}
\label{table:Fukushima1}
\begin{tabular}{l|cccc}
\hline
  & & & $R$ [fm]   &\\
                  &   3   &  5    &  10   &  $ \infty$  \\ 
\hline
$T_{\chi}$ [MeV]  \quad    &  197   & 201   &  203  & 204 \\
$T_{d}$ [MeV]     \quad    &  190   & 192   &  193  & 194 \\
$T_*$ [MeV]       \quad    &  175   & 158   &  137  &   0 \\
\hline
\end{tabular}

\centering
\caption{Same as in Table \ref{table:Polynomial} but for the Polyakov loop potential of Fukushima, version  ${\cal U}_{F2}$.}
\label{table:Fukushima2}
\begin{tabular}{l|cccc}
\hline
  & & & $R$ [fm]   &\\
                    &   3   &  5    &  10   &  $ \infty$  \\ 
\hline
$T_{\chi}$ [MeV] \quad   &  173   & 178   &  181 & 184 \\
$T_{d}$ [MeV]    \quad   &  146   & 146   &  149 & 150 \\
$T_*$ [MeV]      \quad   &  149   & 135   &  118  &   0 \\
\hline
\end{tabular}
\end{table}

Here we will focus on the order parameters for both chiral and deconfinement transitions showing that, as the temperature is increased at zero baryon chemical potential, the PNJL model presents a smooth crossover transition at $T \sim 150 -200$ MeV depending on the size. 
Our results for the bulk are compatible with lattice QCD ones for $N_f = 2+1$, as shown in Ref. \cite{Bazavov:2014} where the authors find a critical temperature of $154 \pm 9$ MeV (see also \cite{Borsanyi:2014}).

The chiral condensate is an order parameter for the spontaneous breaking of chiral symmetry \cite{BJ}. The corresponding crossover transition can be established, for instance, by looking the temperature slope from $\chi_T^{u} \equiv {\partial \langle{\bar{u}}u\rangle}/{\partial T}$ and $\chi_T^{l} \equiv {\partial l}/{\partial T}$. The peak positions give the inflexion points at the chiral critical temperature $T_{\chi}$ and the critical deconfinement temperature $T{_d}$ of the condensates and the Polyakov loop expectation value respectively. As discussed in \cite{Fukushima:2008wg}, it is convenient to take the crossover temperature in the $u-$sector, because the crossover temperature in the $s-$sector is larger and would be far from the deconfinement transition. As also discussed in \cite{Fukushima:2008wg},  the chiral and deconfinement transitions do not take place at the same temperature as long as we treat the chiral condensates and the Polyakov loop as independent variables. Anyway, the idea is exploring different parameters in the Polyakov loop potential to force as much as possible the proximity of both critical temperatures. 
Also, as shown in \cite{Fukushima:2008wg,Ferreira2014,Fu2008} the peak of $\chi_T^{l}$ occurs at a lower temperature than the one of $\chi_T^{u}$, in coincidence with our results presented in Tables $1-4$.

\begin{figure}[tb]
\includegraphics[angle=0,scale=0.37]{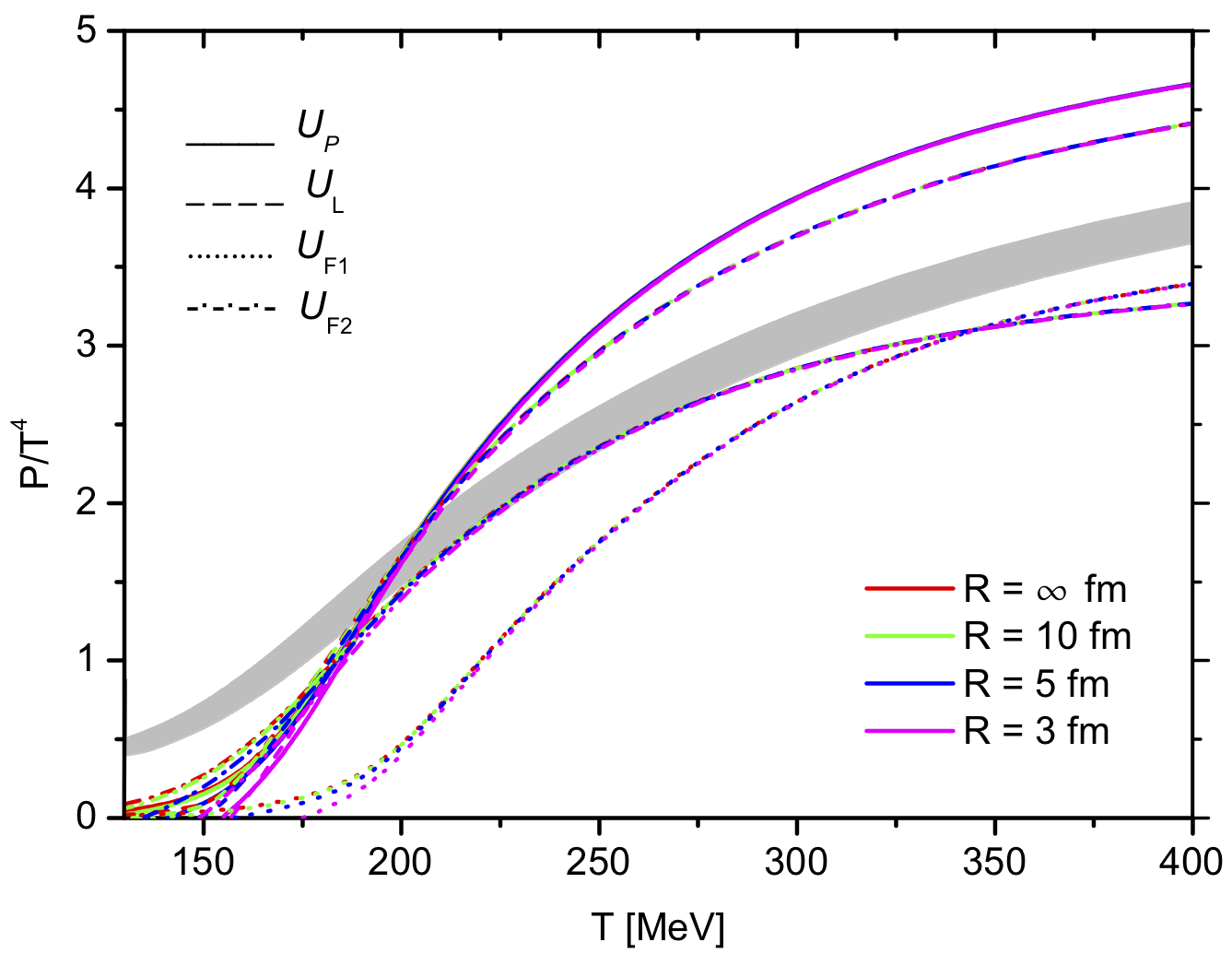}
\caption{We show $P/T^4$ as a function of temperature for different drop sizes and different Polyakov loop potentials. The gray band  are the results for the equation of state in (2+1)-flavor QCD using the highly improved staggered quark action extrapolated to the continuum limit \cite{Bazavov:2014} (see also \cite{Borsanyi:2014}). } 
\label{fig:1}
\end{figure}

For the different choices of the Polyakov loop potential introduced in Sect. \ref{subsec:Polyakov_potentials}, we present in Tables \ref{table:Polynomial}, \ref{table:Logarithmic}, \ref{table:Fukushima1} and \ref{table:Fukushima2} the critical temperatures $T_{\chi}$ and $T{_d}$ for different radii of the system. We also display the temperature $T_*$ below which the drop's pressure $P$ would become negative for the standard prescription of $P_{vac}$.
Below $T_*$ these results would be unphysical  if that $P_{vac}$ is adopted. If another choice of $P_{vac}$ is used the curves in Fig. \ref{fig:1} would shift upwards or downwards and a better coincidence of the model curves with lattice results could be achieved for the bulk case. 
From Tables \ref{table:Polynomial}, \ref{table:Logarithmic}, \ref{table:Fukushima1} and \ref{table:Fukushima2} we see that (except for the cases ${\cal U}_{L}$ and ${\cal U}_{F2}$ with $R=3$ fm) $T_*$ is always below $T_{\chi}$.
Since we are interested in the physics around the critical temperature (relevant for heavy ion collisions and the early universe) we will not show our results for $T < T_*$. 
We emphasize that the values of $T_{\chi}$ and $T{_d}$ are independent of the choice of $P_{vac}$.

\begin{figure*}[tb]
\includegraphics[angle=0,scale=0.37]{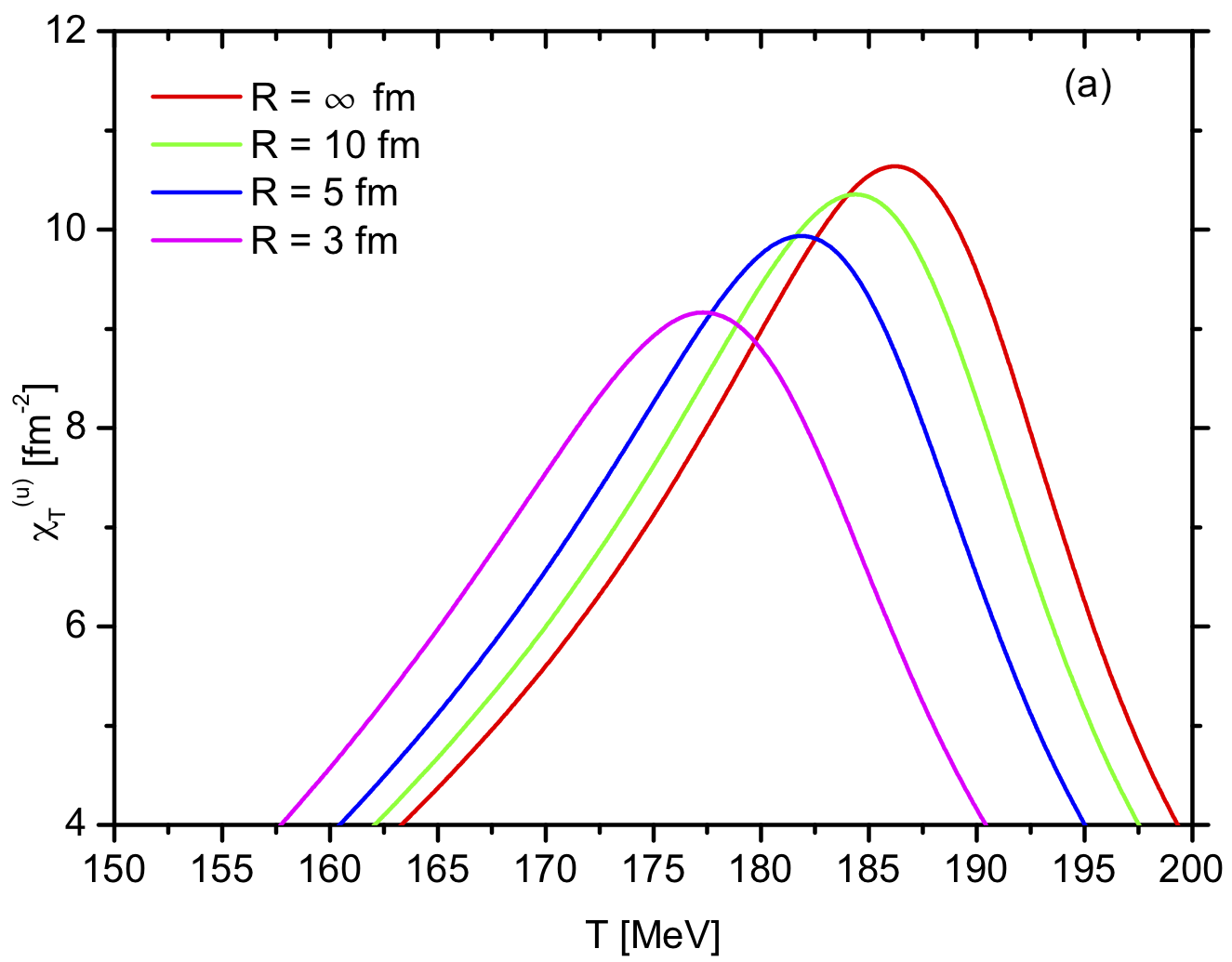}
\includegraphics[angle=0,scale=0.37]{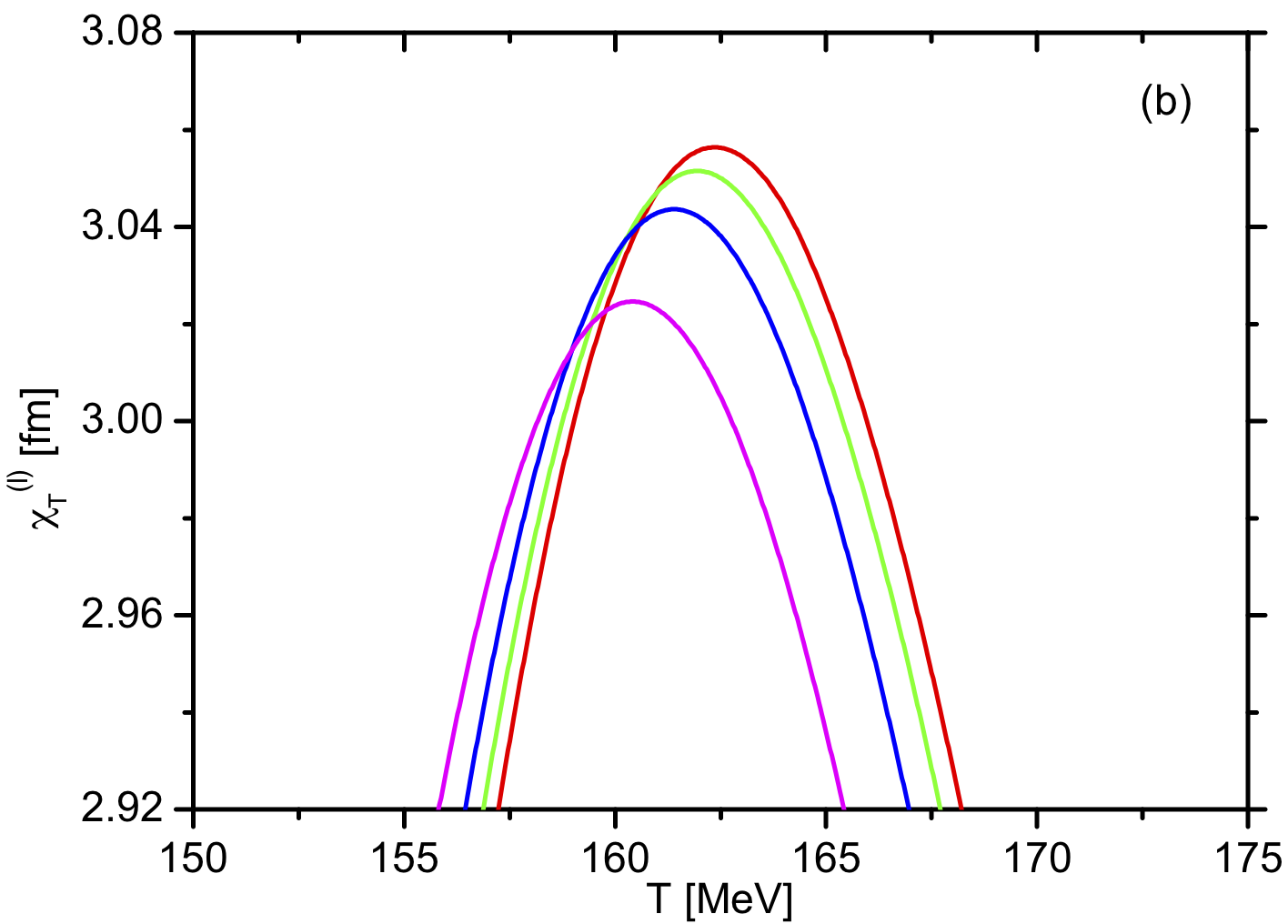}
\caption{We show $\chi_T^{u} \equiv {\partial \langle{\bar{u}}u\rangle}/{\partial T}$ and $\chi_T^{l} \equiv {\partial l}/{\partial T}$ as a function of temperature for the polynomial Polyakov loop potential. }
\label{fig:2}
\end{figure*}

In Table \ref{table:Polynomial}, we show our results for the polynomial Polyakov loop potential. 
The chiral critical temperature $T_{\chi}$ has a significant dependence on the system size; 
it varies from 186 MeV to 177 MeV as the radius shrinks from infinity to 3 fm.  This effect is also apparent in the left panel of Fig. \ref{fig:2} where we see that the peaks of ${\partial \langle{\bar{u}}u \rangle}/{\partial T}$ move towards smaller temperatures as the radius reduces. 
As seen in the right panel of Fig. \ref{fig:2}, ${\partial l}/{\partial T}$ is less sensitive to finite size effects. Thus,  the  critical deconfinement temperature $T{_d}$ varies over a narrower range than $T_{\chi}$, as can be verified in Table \ref{table:Polynomial}. This  behavior could have been anticipated because $l$ feels volume changes only indirectly through the gap equations, and the Polyakov loop potential does not depend explicitly on the size of the system.  As a consequence, $T_{\chi}$ and $T{_d}$ get closer to each other as the drop's size decreases.

\begin{figure*}[tb]
\includegraphics[angle=0,scale=0.37]{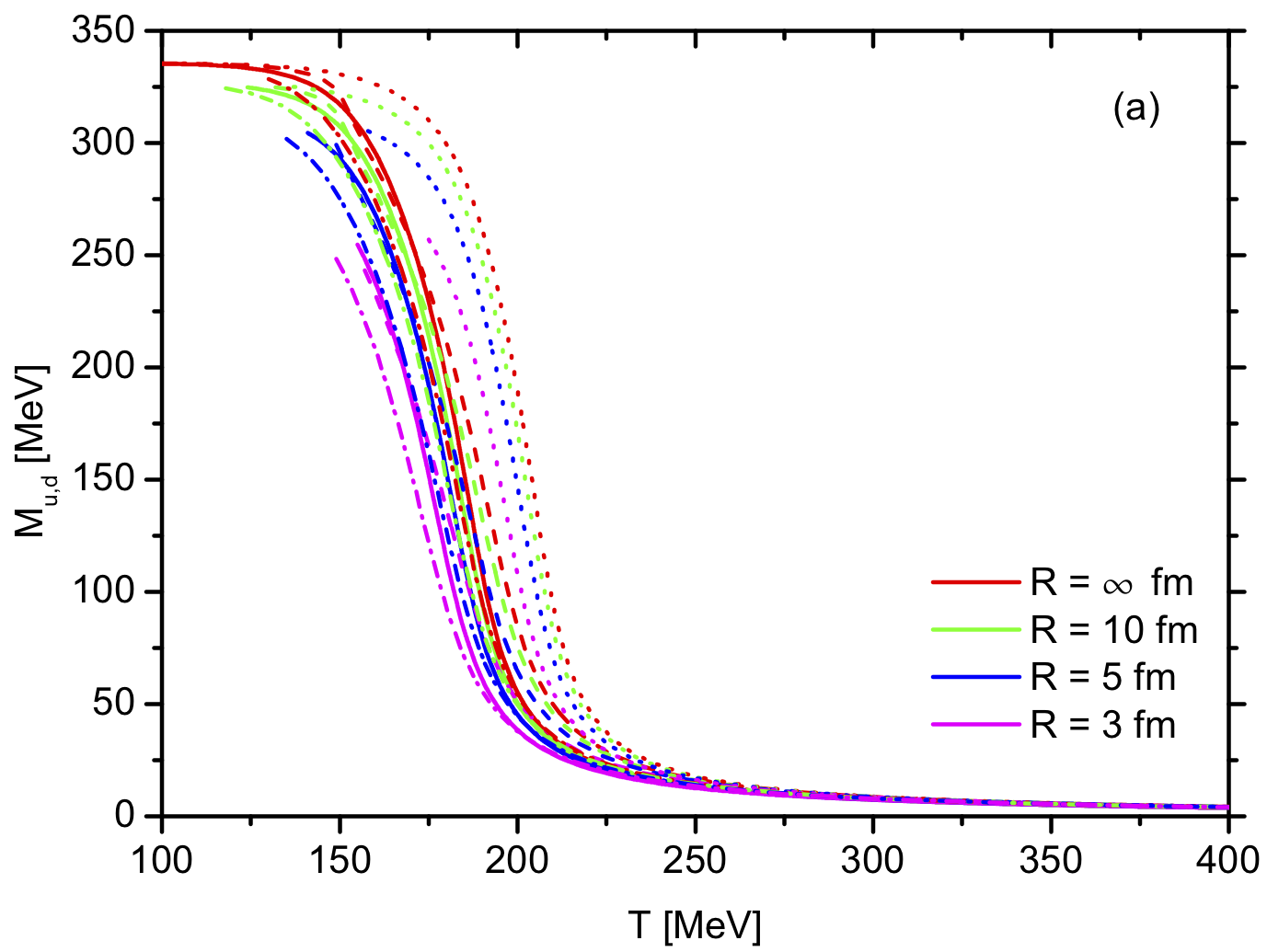}
\includegraphics[angle=0,scale=0.37]{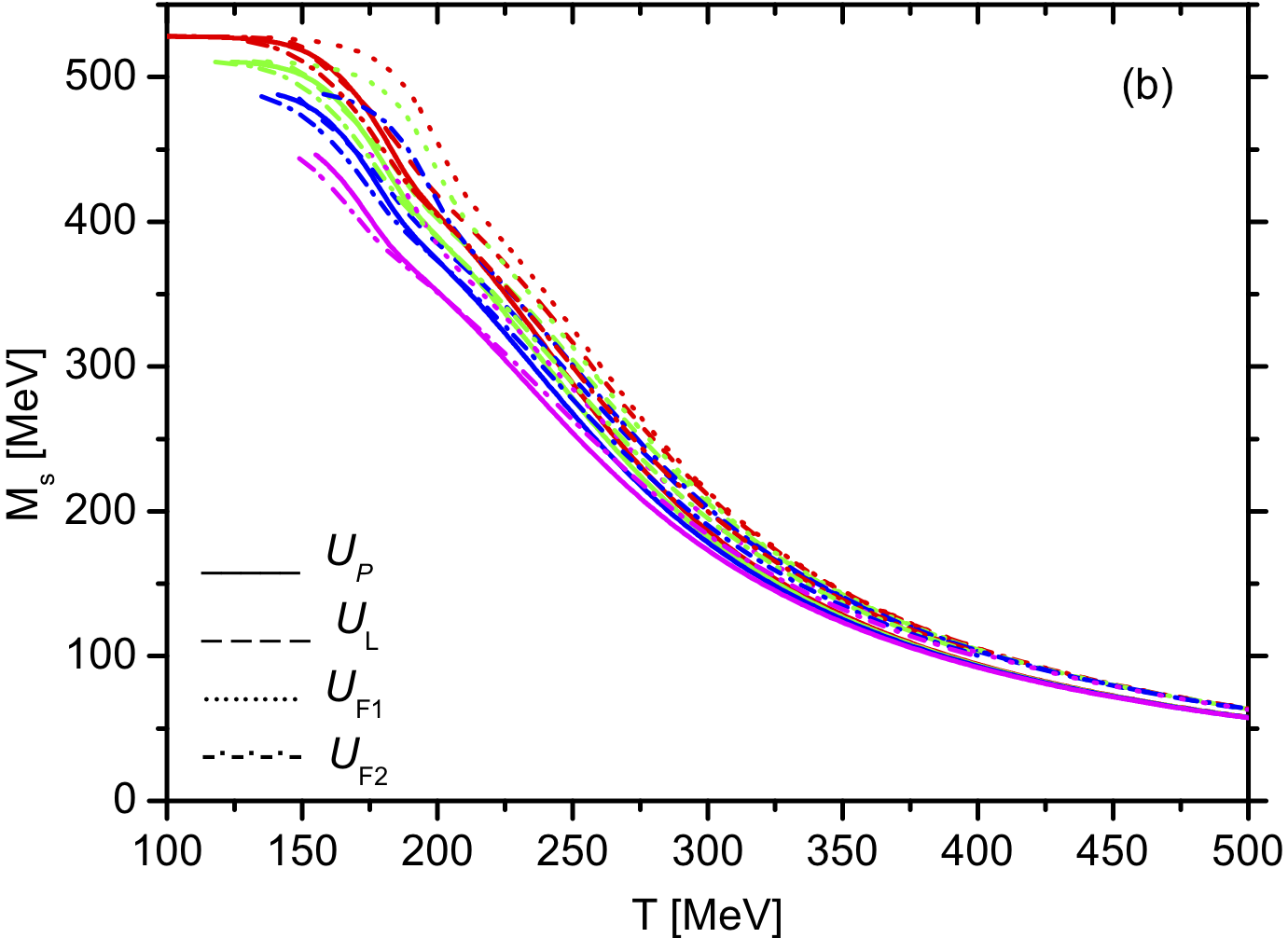}
\caption{Constituent masses $M_u$, $M_d$ and $M_s$ as a function of temperature for different drop sizes and different Polyakov loop potentials. We do not show the branch of each curve corresponding to temperatures for which the drop's pressure becomes negative for the standard choice of $P_{vac}$.} 
\label{fig:3}
\end{figure*}

The critical temperatures for the model with a logarithmic Polyakov loop potential can be seen in Table \ref{table:Logarithmic}. In this case $T_{\chi}$ varies from 192 MeV to 181 MeV as $R$ decreases. 
Here the deconfinement temperatures are slightly smaller than in the previous case, and the chiral ones, larger. For $R = 3$ fm, we find that $T{_d}$ lies in the negative pressure region for the standard choice of $P_{vac}$.

In the cases with ${\cal U}_L$ and ${\cal U}_P$, the choice of $T_0$ affects both, the deconfinement and the chiral critical temperatures. Here we use  $T_0 = 185$ MeV in agreement with the values used in \cite{Bhattacharyya2013,Schaefer:2007pw,BJ} for $N_f=2+1$. For larger $T_0$, $T_{\chi}$ and $T_d$ approach each other but both values increase, spoiling the coincidence with lattice results.  On the other hand, for smaller $T_0$, $T_{\chi}$ and $T_d$ are closer to lattice data but there is larger separation between them.

Finally, we show the critical temperatures for the Polyakov loop potential of Fukushima. Here we considered two different examples, as discussed in the previous section. In Table \ref{table:Fukushima1} we show the results for  $b = (196.2 \, \mathrm{MeV})^3$, and in Table \ref{table:Fukushima2} for $b = (115 \, \mathrm{MeV})^3$. 
The first case gives higher $T_{\chi}$ and $T_d$ but both temperatures are closer to each other. In the second case we obtain smaller critical temperatures (closer to lattice results for 2+1 flavors) but there is a larger separation between them. 

Summing up, in all cases discussed above, we see as a common behavior that $T_{\chi}$ decreases with the size of the system  by around $5 \% $ when the radius goes from $R = \infty$ to $R = 3$ fm. We also note that $T_d$ varies by less than $2 \% $ in the same size interval.
As a consequence, as the drop's size decreases, $T_{\chi}$ becomes closer to $T{_d}$. This behavior is in agreement with the results presented in \cite{Cristoforetti2010,Bhattacharyya2013}.

In Fig. \ref{fig:3} we show the temperature variation of the constituent masses $M_u$, $M_d$ and $M_s$ for different drop sizes and for all the Polyakov loop potentials presented in Sect. \ref{subsec:Polyakov_potentials}.
In the chirally broken phase, we find that the constituent quark masses are somewhat smaller for drops with smaller radii. In this region, the volume dependence of the effective masses is stronger than in the chirally restored phase. Also, $M_u$ and $M_d$ show a steep slope around the crossover temperature while for $M_s$ the slope is smoother. 
As shown in  in Tables \ref{table:Polynomial}$-$\ref{table:Fukushima2}, the chiral critical temperature $T_{\chi}$ shifts to smaller values as the volume decreases. 
Such behavior is also apparent in Fig. \ref{fig:3} where we see that, for smaller systems, the constituent mass tends to the current value at lower temperatures. A similar behavior has been reported in \cite{Bhattacharyya2013}.

\subsection{Interaction measure}
\label{sec:interaction_measure}

\begin{figure*}[tb]
\includegraphics[angle=0,scale=0.37]{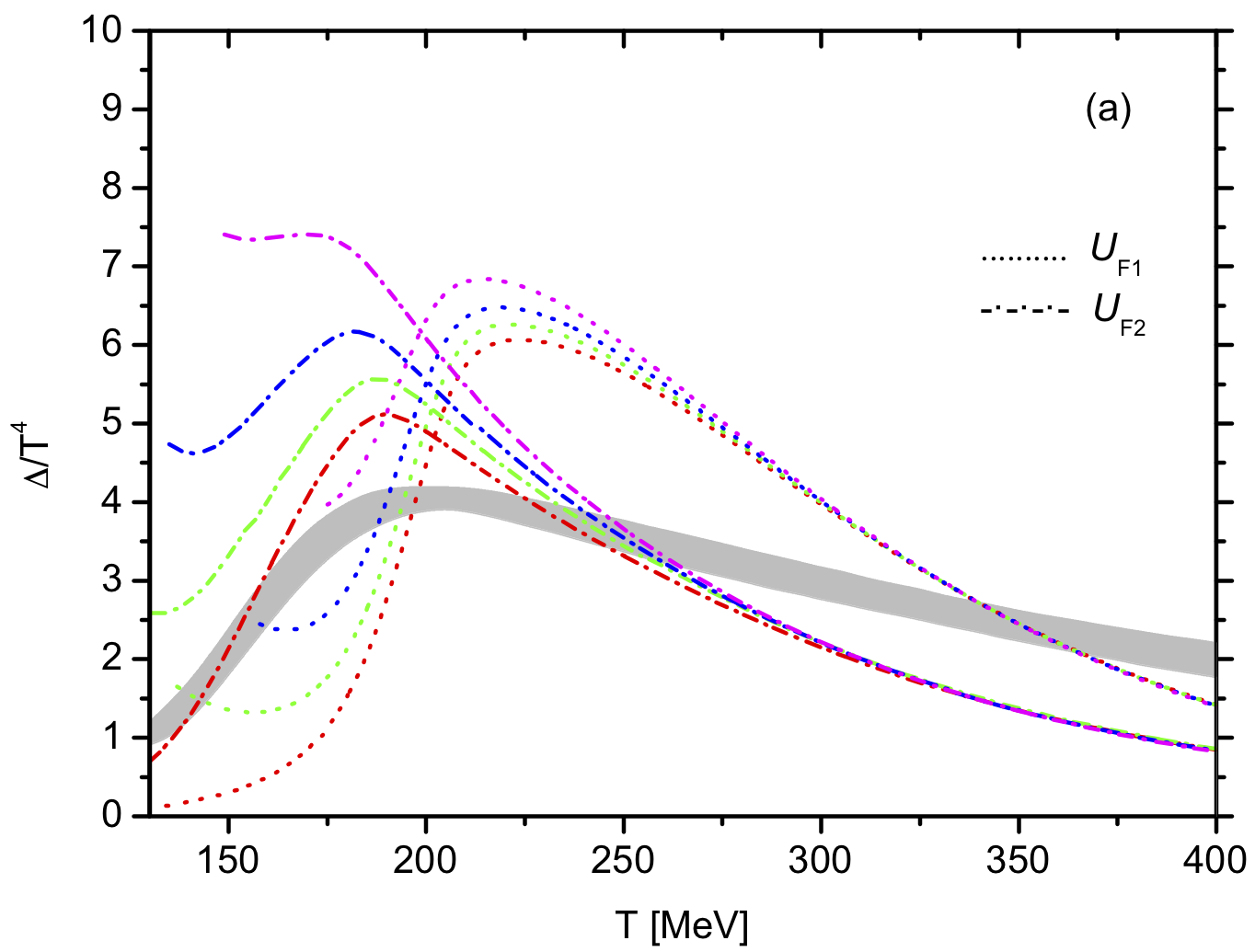}
\includegraphics[angle=0,scale=0.37]{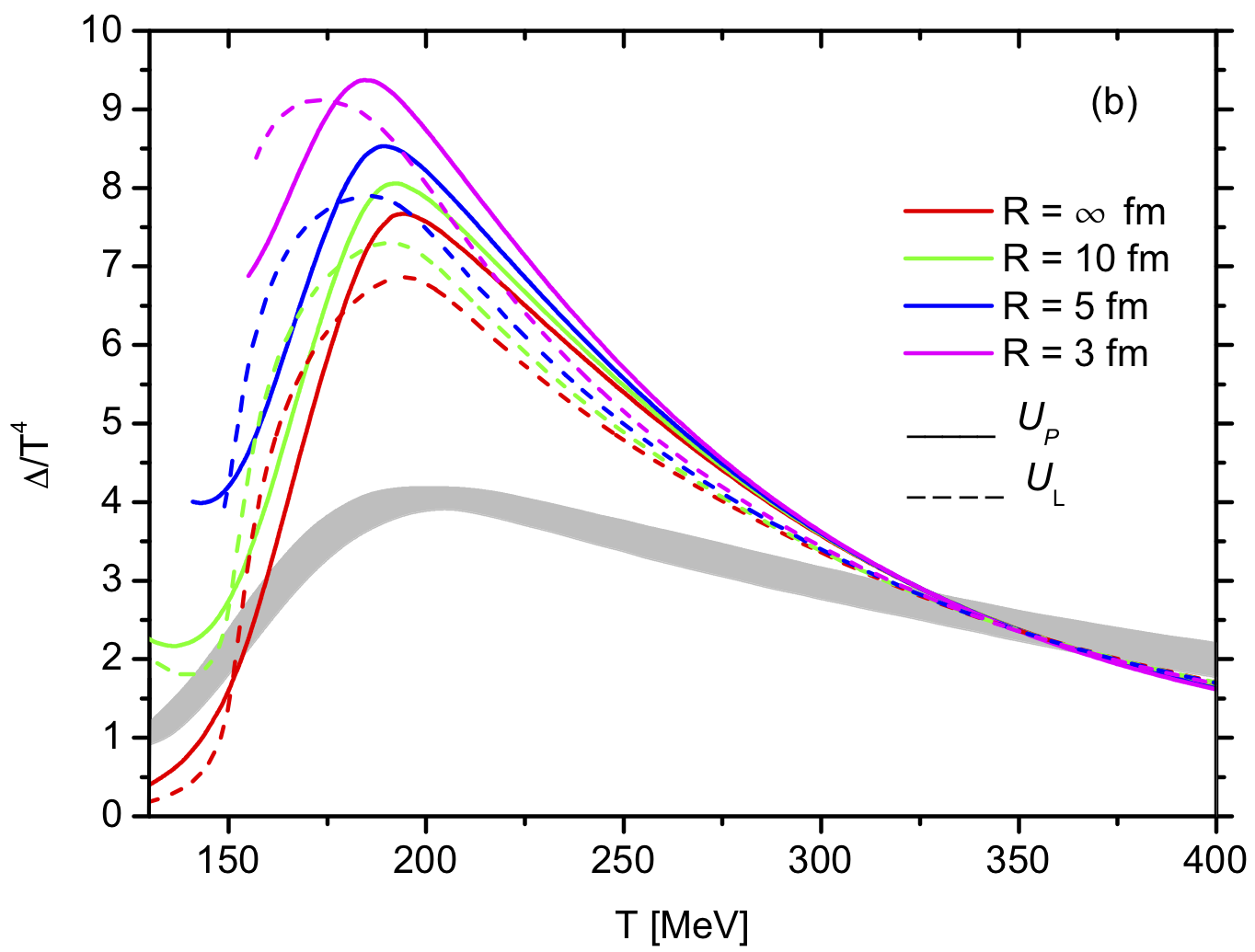}
\caption{We show $\Delta/T^4$ as a function of temperature for different drop sizes and different Polyakov loop potentials. We also include lattice QCD simulations data from \cite{Bazavov:2014} (gray band).}
\label{fig:4}
\end{figure*}

A thermodynamic quantity of special interest is the thermal expectation value of the trace of the energy momentum tensor:
\begin{eqnarray}
\Theta^{\mu\mu} (T) \equiv  \epsilon(T)  - 3 P(T)  .
\end{eqnarray}
This quantity  is known as trace anomaly  or equivalently as \textit{interaction measure} $\Delta(T) \equiv \epsilon(T)  - 3 P(T)$  since it is very sensitive to the non-perturbative effects in the quark-gluon plasma. 
Specifically,  it measures the deviation from the equation of state of an ideal gas $\epsilon = 3 P$ due to interactions and/or finite quark masses. 

Here we focus on the quantity 
\begin{eqnarray}
\frac{\Delta(T)}{T^4} = \frac{ \epsilon(T)  - 3 P(T)}{T^4}  
\end{eqnarray}
which allows a straightforward assessment of deviations from the Stefan-Boltzmann behavior.

Within the present model, the energy density $\epsilon(T)$ at zero chemical potential is given by
\begin{eqnarray}
\epsilon(T) = \frac{\mathbf{\Omega}_{\mathrm{MRE}}(T)}{V} + T s(T)
\end{eqnarray}
where the entropy density is given by:
\begin{eqnarray}
s(T) = -   \frac{1}{V}  \frac{\partial \mathbf{\Omega}_{\mathrm{MRE}}(T)}{\partial T} 
\end{eqnarray}

The interaction measure is sensitive to the finite drop's volume, because the energy density has an explicit dependence on the surface tension and the curvature energy:
\begin{equation}
\epsilon(T)= - P(T) + \alpha(T) \; \frac{S}{V} + \gamma(T) \; \frac{C}{V} + T \; s(T),
\end{equation}
In addition, as apparent from Eqs. \eqref{P}, \eqref{surfacetension} and \eqref{curvatureenergy},  there is an additional dependence on finite  size effects through the infrared cutoff $\Lambda_{i, \mathrm{IR}}$ in the integrals for $P$, $\alpha$ and $\gamma$.

In Fig. \ref{fig:4} we show our results for the bulk and for finite size systems together with lattice QCD simulation data in the continuum limit \cite{Bazavov:2014}.  
In general, we observe that the predictions of our effective model in bulk are in qualitative agreement with lattice QCD results. The peak heights are somewhat larger that in lattice QCD; nonetheless, the peak positions are in good coincidence with lattice.
As a global feature, common to all finite sizes models that include different Polyakov loop potentials $\cal U$, the interaction measure presents a peak that moves towards decreasing temperatures as the radii decrease.  
Note that, even though the interaction measure is explicitly dependent on $P_{vac}$, the temperatures at which the peaks take place are not affected by the $P_{vac}$ choice.  

For the chirally broken phase, i.e. for temperatures below the one in the peak, the curves for the bulk case are in qualitative agreement with lattice data. 
Close to $T_*$, for finite sizes, the curves have a local minimum and start to increase at lower temperatures due to the contribution of the surface tension and the curvature energy.

Now let us concentrate on the peak of the curves. For $R = \infty$, the peak position of the curves  with ${\cal U}_L$, ${\cal U}_P$ and ${\cal U}_{F2}$ are in better coincidence with lattice results. The one with ${\cal U}_{F1}$ is shifted to higher temperatures. For finite systems the position of the peaks is shifted to lower temperatures in all cases. 
As a global feature, the peak heights with $R = \infty$ for all models are larger than lattice results. We get a better agreement for ${\cal U}_{F2}$ which is a $\sim 25\%$ higher than lattice.  For finite systems we see that the height of the peaks increase as the radii decrease, and they shift to smaller temperatures.

For high enough temperatures, in the chirally restored phase,  there is a reasonable agreement between the bulk models and lattice results, specially for the ${\cal U}_L$ and ${\cal U}_P$ Polyakov potentials. Results for the ${\cal U}_{F1}$ and ${\cal U}_{F2}$ potentials, are somewhat below the lattice data. For finite sizes our results are superposed with the corresponding bulk case. 

\subsection{Energy density and entropy density}

\begin{figure*}[tb]
\includegraphics[angle=0,scale=0.37]{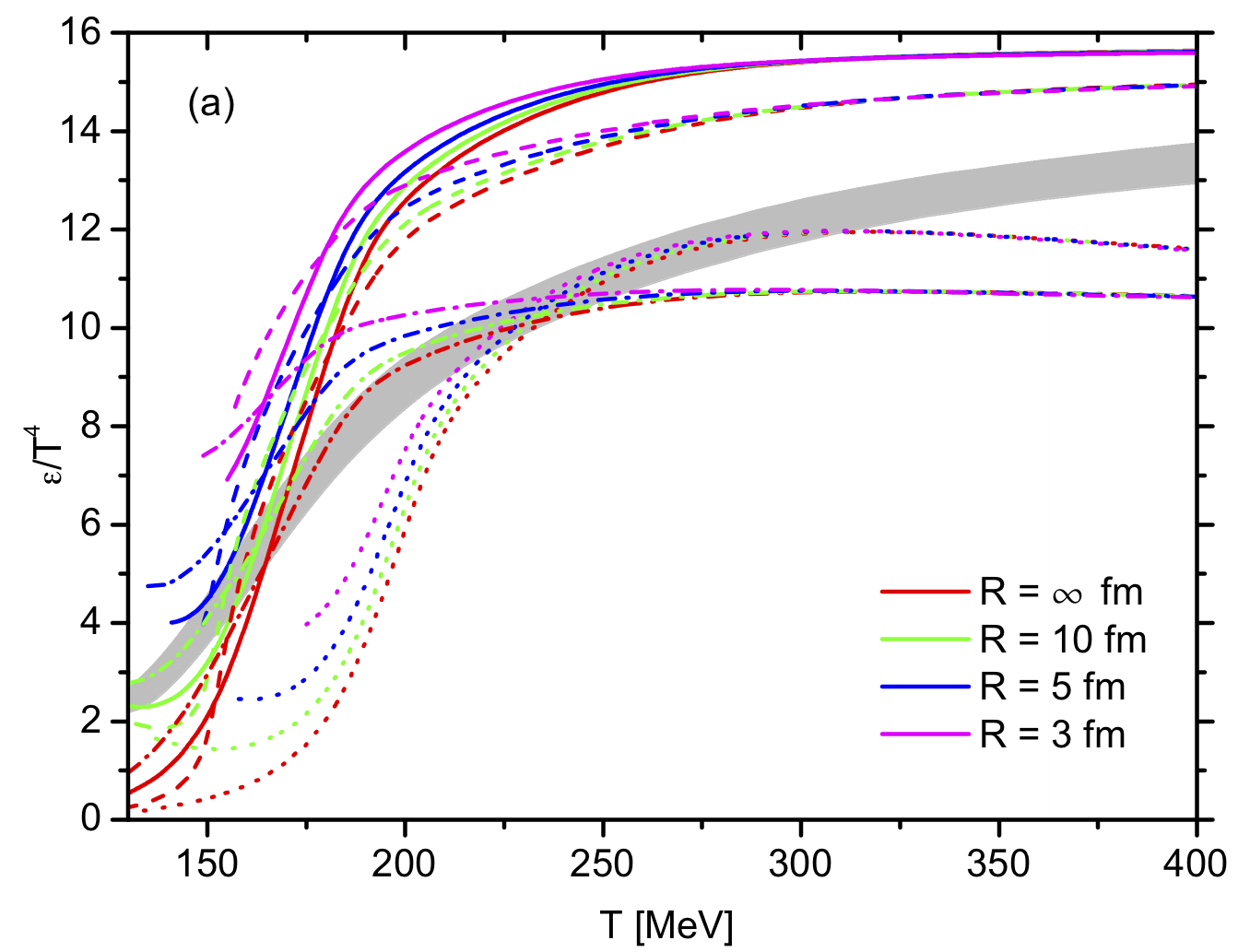}
\includegraphics[angle=0,scale=0.37]{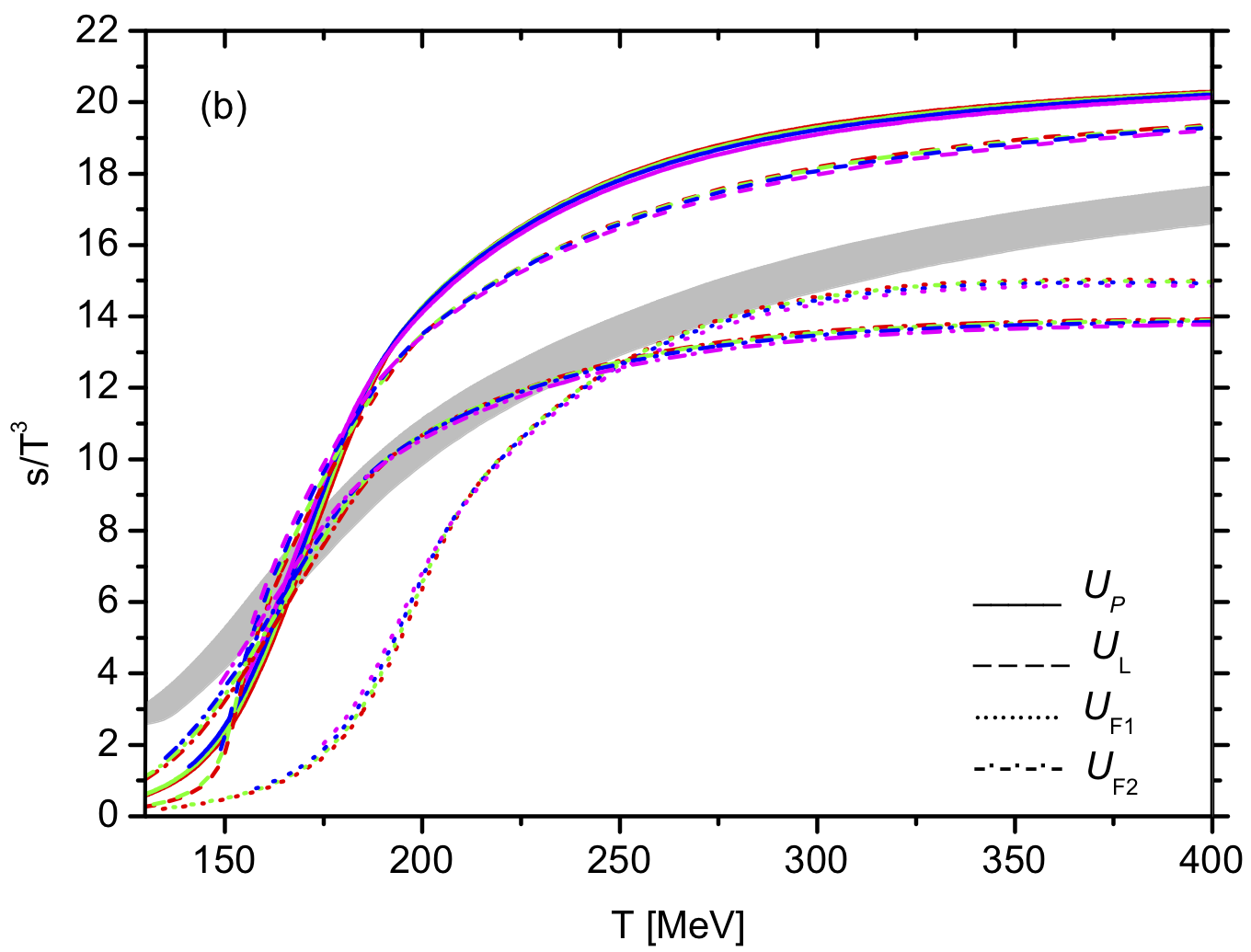}
\caption{Energy density and entropy density as functions of temperature. Lattice data (gray band) are taken from \cite{Bazavov:2014}.}
\label{fig:5}
\end{figure*}

In the bulk case, our results for the energy density and the entropy density are in qualitative agreement with lattice QCD results (see Fig. \ref{fig:5}) and with Ref. \cite{BJ}, as can be seen from their Fig. 3. 
As previously mentioned for the results of Fig. \ref{fig:1},  different choices of $P_{vac}$ would lead only to a vertical shift of the curves for the energy density but will not change the temperature of the inflexion points. One could take advantage from this feature and introduce a different procedure to fix $P_{vac}$ in such a way that our predictions for the bulk case are as close as possible to lattice data. Since our focus here is not centered on the equation of state we shall not explore such strategy in the present work.

For high enough temperatures, our curves for all thermodynamic quantities  approach to the Stefan-Boltz\-mann limit. The Stefan-Boltzmann limit for the pressure  is given by
\begin{eqnarray}
\frac{p_{SB}}{T^4} = (N_c^2 -1) \frac{\pi^2}{45} +  N_c N_f \frac{7 \pi^2}{180} ,
\end{eqnarray}
where $N_c$ and $N_f$ are the number of colors and flavors. The first term represents the gluonic contribution and the second, the quark's contribution. For  $N_c=3$ and $N_f=3$ we have: 
\begin{equation}
\frac{p_{SB}}{T^4} =  \frac{8\pi^2}{45} +  \frac{7 \pi^2}{20} = 1.75 + 3.45 = 5.20
\end{equation}
which results in  $\epsilon_{SB}/T^4 = 5.26 + 10.36 = 15.62$ and  $s_{SB}/T^3 = 7 + 13.8 = 20.8$.

From Fig. \ref{fig:5} we see that, at high enough temperatures,  models with the Fukushima potentials ${\cal U}_{F1}$ and  ${\cal U}_{F2}$ tend to the  Stefan-Boltzmann limit for quarks only (no gluons) while models with the potentials ${\cal U}_{P}$ and ${\cal U}_{L}$ tend to the  Stefan-Boltzmann limit  including quarks and gluons.

This behavior is already known from previous works \cite{Fukushima:2008wg,BJ,Bhattacharyya2013}.
In the case of the ${\cal U}_{P}$ and ${\cal U}_{L}$ potentials, both the unconfined transverse gluons as well as the Polyakov loop, which corresponds to longitudinal gluons, contribute to the thermodynamic quantities \cite{Fukushima:2008wg}. 
But, since the Polyakov-loop potentials are fitted to pure gauge lattice data, they thus reproduce the total pressure, energy density, and entropy density of both the longitudinal and the transverse gluons, overcounting the degrees of freedom in the chirally symmetric phase \cite{Fukushima:2008wg,BJ}. However, the potential ansatz by Fukushima excludes these transverse gluon contributions at high temperatures leading to the differences found in Fig. \ref{fig:5}. Nonetheless, at temperatures around and below the transition temperature such differences tend to disappear. 

It is worth to remark that in Fig. \ref{fig:5} there is a wide range of temperatures in which our results for ${\cal U}_{F2}$ are in a quantitatively good agreement with lattice results.

For finite systems, we see that in all cases the curves converge to the bulk ones at  high temperatures. Close to the transition region, the curves for different radii start to separate each other as the temperature decreases. In coincidence with Ref. \cite{Bhattacharyya2013}, we find that the smaller the radius the higher the temperature at the inflexion point. Nonetheless, in Ref. \cite{Bhattacharyya2013} the results for $R = 5$ fm and for the bulk case are coincident for all temperatures but in our case are not.  In the chirally broken phase,   the energy density and the entropy density change very little with the drop's size.

\subsection{Specific heat and  speed of sound}

\begin{figure*}[tb]
\includegraphics[angle=0,scale=0.37]{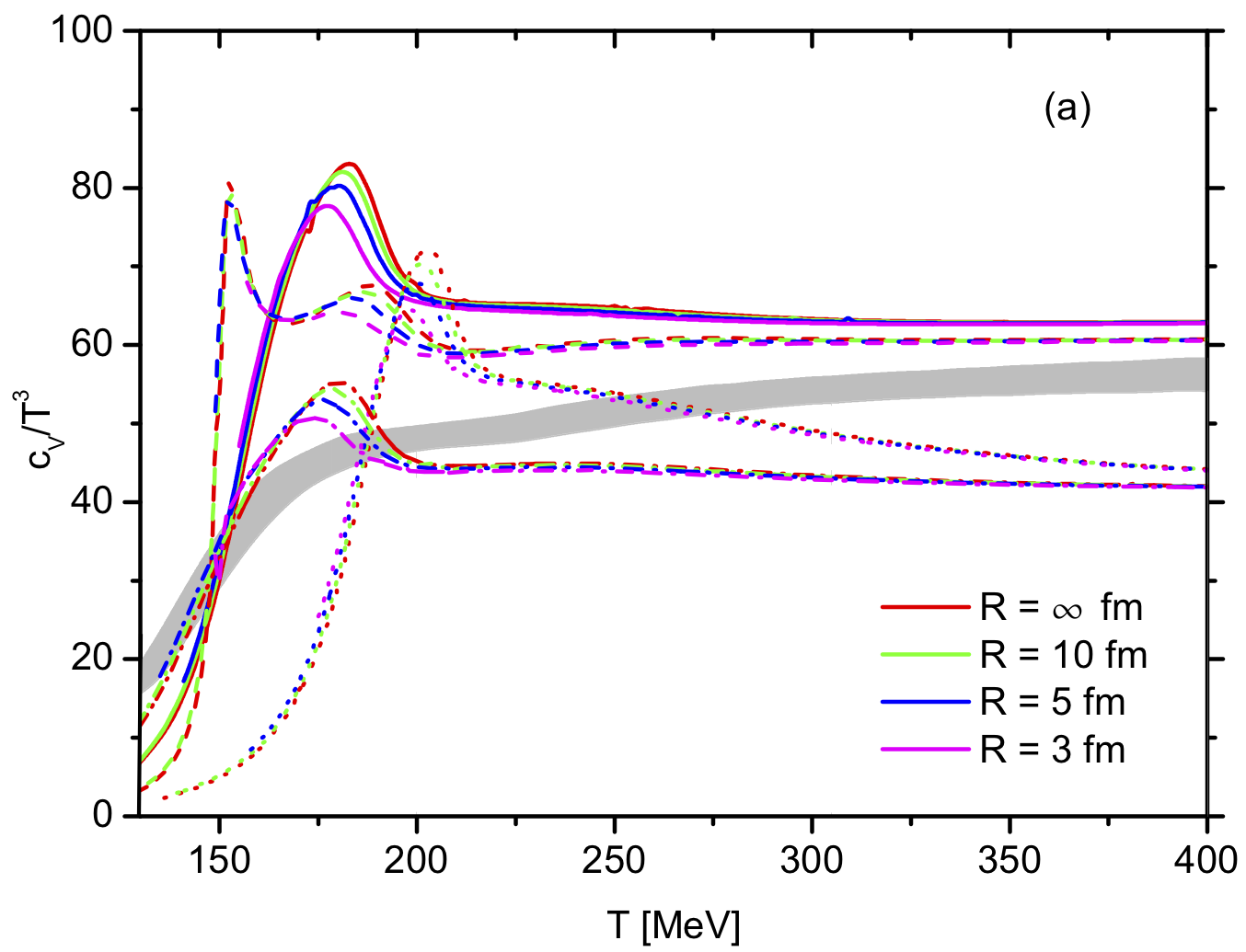}
\includegraphics[angle=0,scale=0.37]{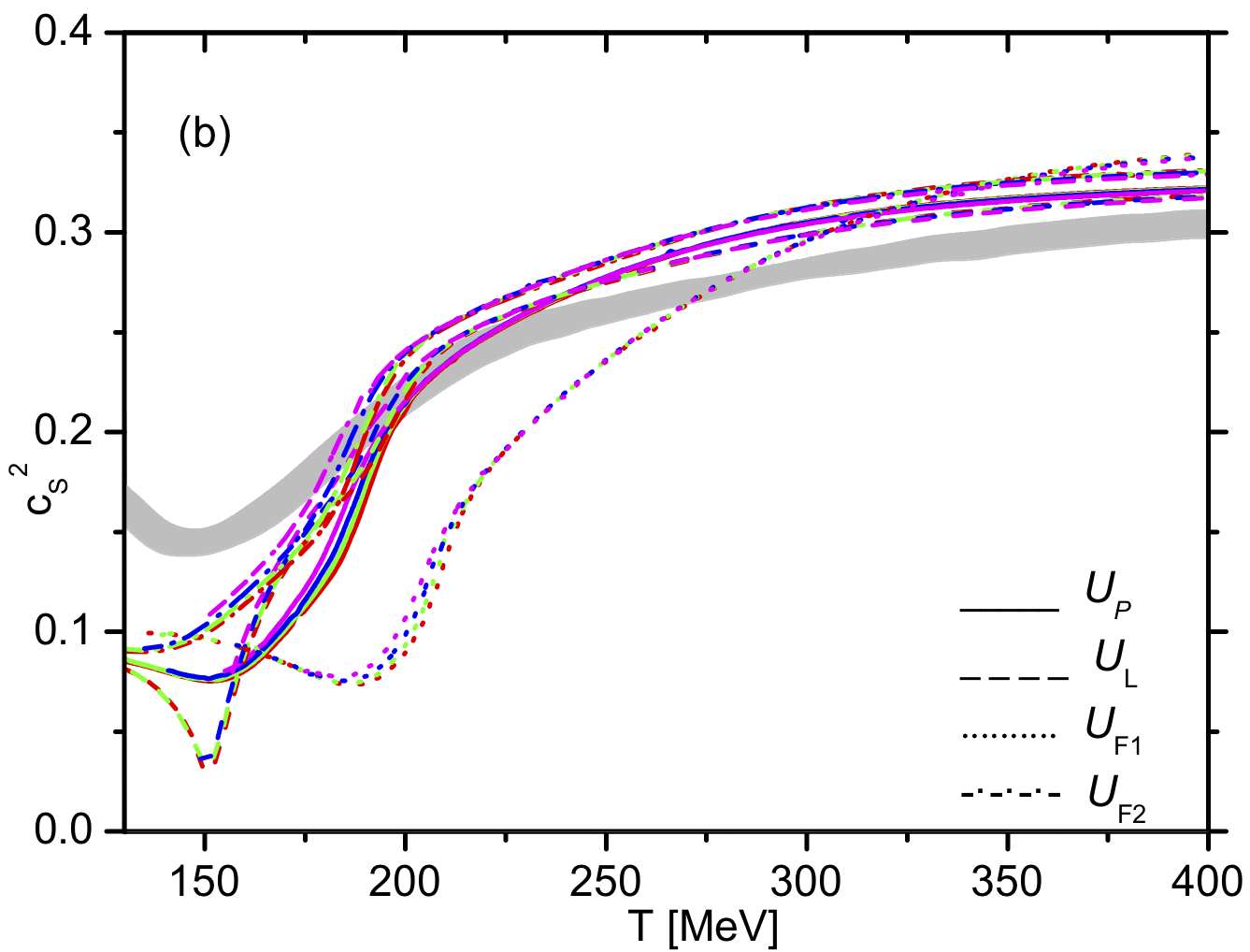}
\caption{Specific heat and speed of sound. Lattice data (gray band) are taken from \cite{Bazavov:2014}.}
\label{fig:6}
\end{figure*}

The specific heat at constant volume is given by
\begin{equation}
c_V = -T \frac{\partial^2 \mathbf{\Omega}_{\mathrm{MRE}}}{\partial T^2} \bigg|_V,
\end{equation}
and the corresponding results are summarized in Fig. \ref{fig:6}.   At low temperatures $c_V$ grows with $T$, then
shows a peak at the transition temperature, and approaches the corresponding Stefan-Boltzmann limit  for high enough $T$. 
For the bulk case our results are in agreement with  Ref. \cite{BJ} and with lattice data \cite{Bazavov:2014}. In fact, in the left panel of Fig. \ref{fig:6} we note that lattice data show a soft undulation around the critical temperature, whose position is close to the peaks considering ${\cal U}_{F2}$ and  ${\cal U}_{P}$. For ${\cal U}_{F1}$ and  ${\cal U}_{L}$ the peaks are shifted to higher temperatures. In general, the best agreement with lattice data (up to the critical temperature and somewhat above it) is obtained with ${\cal U}_{F2}$.

For finite size drops, we find that $c_V$ doesn't change significantly  with the change in volume, except in the crossover region. In fact, we  find that  the height of the peaks decreases as the volume shrinks,  in agreement with \cite{Bhattacharyya2013}.  Also, the peak position shifts to smaller temperatures as the volume decreases, as in  \cite{Bhattacharyya2013}. 
As for other thermodynamic quantities, we find that the specific heat for the models with ${\cal U}_{F1}$ and ${\cal U}_{F2}$ tend to the Stefan-Boltzmann limit for quarks while the models with ${\cal U}_{P}$  and  ${\cal U}_{L}$ tend  to the Stefan-Boltzmann limit for quarks and gluons,  due to the differences in the contributing gluon degrees of freedom.

In Fig. \ref{fig:6} we show our results for the speed of sound \cite{BJ}
\begin{equation}
c_s^2 = \frac{\partial p}{\partial \epsilon} \bigg|_S  = \frac{s}{c_V}. 
\end{equation}
The behavior of $c_s^2$ is associated directly with the role of interactions in the system. The strength of interactions can be quantified through the interaction measure $\Delta$ calculated in Sec. \ref{sec:interaction_measure}. A comparison between $\Delta$ presented in Fig. \ref{fig:4} and  $c_s^2$ depicted in Fig. \ref{fig:6}, shows that these quantities are correlated. At large temperatures, as the value of $\Delta$ goes to zero, the speed of sound tends to the ultarelativistic limit of an ideal gas, $c_s = 1/\sqrt{3}$. At lower temperatures, interactions become relevant and therefore $\Delta$ grows and $c_s^2$ decreases significantly.

Except for ${\cal U}_{F1}$, all minima positions lie close to the lattice QCD one. In the chirally restored phase our results for all Polyakov potentials are in good agreement with lattice QCD data, except for the ${\cal U}_{F1}$ case that approaches lattice at higher temperatures.

Contrary to previous findings \cite{Bhattacharyya2013}, our results show that the speed of sound doesn't depend too much on the system's size. In fact, small variations are observed only in the transition region.

\subsection{Surface tension and curvature energy}

\begin{figure*}[tb]
\includegraphics[angle=0,scale=0.37]{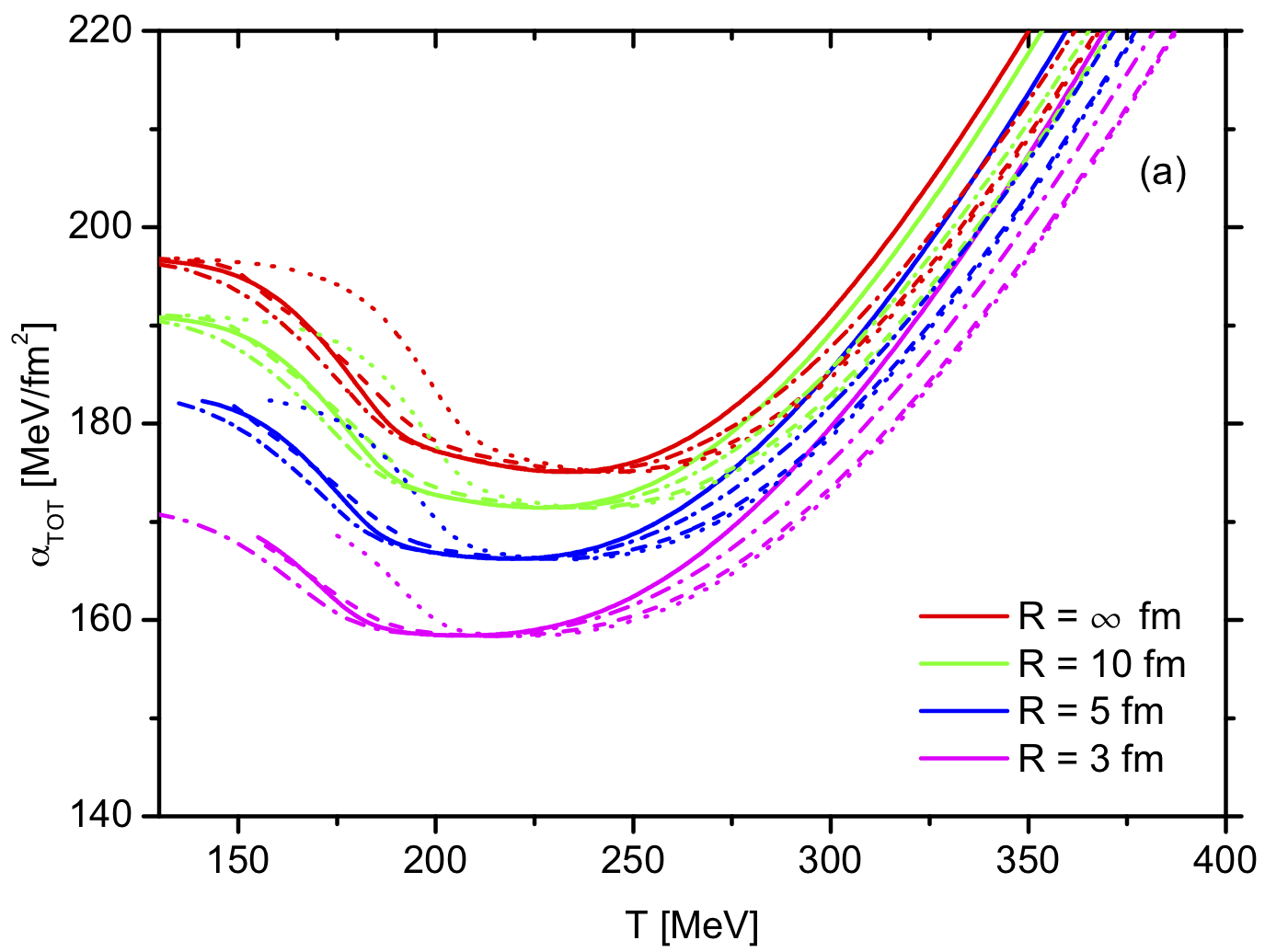}
\includegraphics[angle=0,scale=0.37]{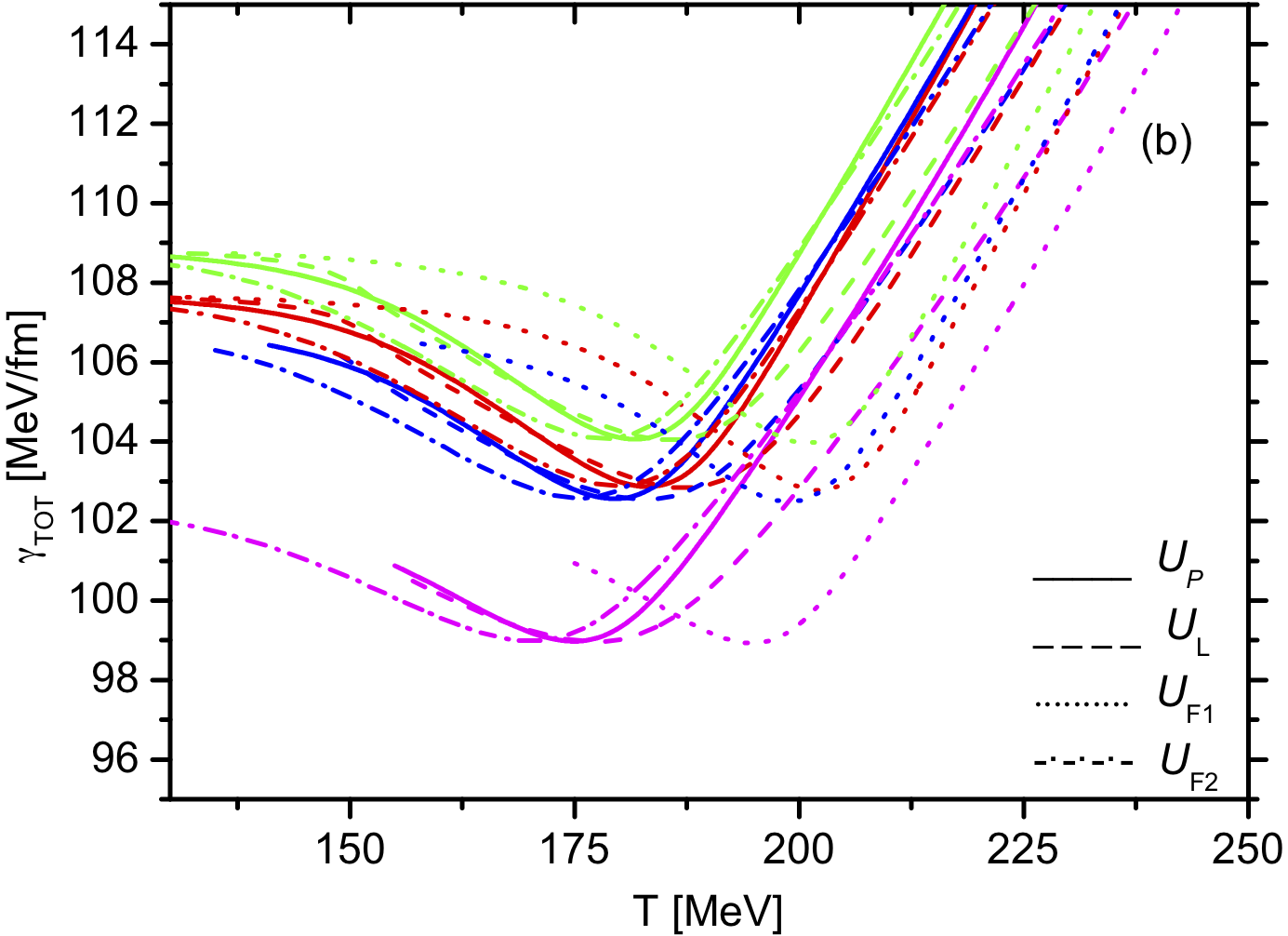}
\caption{Total surface tension and total curvature energy as a function of temperature; the scales of panels (a) and (b) are different. The minimums of  $\alpha_{\mathrm{TOT}}$ and of $\gamma_{\mathrm{TOT}}$ occur at different temperatures.}
\label{fig:7}
\end{figure*}

In Fig. \ref{fig:7} we show the total surface tension $\alpha_{\mathrm{TOT}}$ and the total curvature energy $\gamma_{\mathrm{TOT}}$ for drops with different sizes. $\alpha_{\mathrm{TOT}}= \sum_i \alpha_i$ and $\gamma_{\mathrm{TOT}}= \sum_i \gamma_i$ include the contribution of $u$, $d$ and $s$ quarks.  
We have checked that $\alpha_{s}$ is more than $10$ times larger than $\alpha_{u}$ and $\alpha_{d}$, in qualitative accordance with results for cold quark matter at very high densities \cite{Lugones:2013bo,Lugones:2017ft} that show that the  total surface tension $\alpha_{\mathrm{TOT}}$ is largely dominated by the contribution of strange quarks. On the contrary, $\gamma_{u}$ and $\gamma_{d}$ are typically $\sim 1-2$ times $\gamma_{s}$ and thus the behavior of $\gamma_{\mathrm{TOT}}$ is controlled mainly by $u$ and $d$ quarks.

Both, $\alpha_{\mathrm{TOT}}$ and $\gamma_{\mathrm{TOT}}$ show a significant variation with $R$ at all temperatures, specially for small drops with radii below 10 fm. 
There is also a considerable dependence on the Polyakov loop potential. 

At large temperatures $\alpha_{\mathrm{TOT}}$ and $\gamma_{\mathrm{TOT}}$ are monotonically increasing functions of $T$. Moreover, for $T \gtrsim 250$ MeV  the surface tension grows approximately as
\begin{equation}
\alpha_{\mathrm{TOT}} = C_{\alpha} T^{3/2},
\end{equation}
being $C_{\alpha} \approx 0.029-0.034 \, \mathrm{MeV}^{-1/2} \mathrm{fm}^{-2}$,  while the curvature energy grows as 
\begin{equation}
\gamma_{\mathrm{TOT}} = C_{\gamma} T^{3/2},
\end{equation}
being $C_{\gamma} \approx 0.030-0.035 \, \mathrm{MeV}^{-1/2} \mathrm{fm}^{-1}$.

At lower temperatures both $\alpha_{\mathrm{TOT}}$ and $\gamma_{\mathrm{TOT}}$ have local minimums. In the case of the total curvature energy the minimum falls around the chiral critical temperature $T_{\chi}$ of the $u$ and $d$ condensates, which evidences the fact that $\gamma_{\mathrm{TOT}}$ is  controlled mostly by up and down quarks and is sensitive to their chiral transition. On the other hand, the total surface tension is sensitive to the chiral transition of strange quarks and therefore its minimum falls at a larger temperature. 

At temperatures below that of the minimum there is a narrow interval where  $\alpha_{\mathrm{TOT}}$ and $\gamma_{\mathrm{TOT}}$ are decreasing functions of $T$. For even smaller temperatures,  $\alpha_{\mathrm{TOT}}$ and $\gamma_{\mathrm{TOT}}$ tend to a constant value which is of the same order of the values obtained within the NJL model  for cold quark matter ($T=0$) at finite chemical potentials ($\mu = 0-450$ MeV) \cite{Lugones:2013bo}. 
In some cases such constant value is not shown in the figures because the pressure becomes negative for the standard choice of $P_{vac}$.

\section{Summary and conclusions}
\label{sec:conclusions}
In this work we studied the thermodynamic properties of finite systems composed by quark matter containing two light and one heavy quark within the frame of the PNJL model. We have considered vanishing baryon chemical potential and finite temperatures. 
We compared our numerical results for the bulk case with those from lattice QCD simulations, and then we studied the finite size deviations from the bulk case. 
We  included finite size effects through the Multiple Reflection Expansion formalism and explored the effect of using different Polyakov loop potentials. 
Finite size effects were incorporated in the fermion integrals but not in the Polyakov loop potentials. {However, if the pure Yang-Mills theory were formulated with a finite radius, the  deconfinement phase transition could be affected and presumably the first-order transition would turn into a smooth crossover for small enough radii.} This is beyond the scope of the present work. 

As the temperature is increased at zero baryon chemical potential, the order parameters for both chiral and deconfinement transitions indicate that the PNJL model presents a smooth crossover transition, in accordance with lattice QCD results.  
For different radii of the system and different choices of the Polyakov loop potential, we determined the chiral critical temperature $T_{\chi}$ of the $u$ and $d$ condensates and the critical deconfinement temperature $T{_d}$ of the Polyakov loop expectation value  (see Table \ref{table:Polynomial}). 
In general, $T_{\chi}$ depends on the system's size, decreasing {by around $5 \% $ when the radius goes from infinity to 3 fm, while $T{_d}$ varies by less than $2 \%$ in the same interval.} Thus, as the drop's size decreases, $T_{\chi}$ becomes closer to $T{_d}$, in accordance with \cite{Cristoforetti2010,Bhattacharyya2013}.

Then we focused on the interaction measure $\Delta(T) \equiv \epsilon(T)  - 3 P(T)$,  which evaluates the deviation from an ideal gas behavior ($\epsilon = 3 P$) due to interactions and/or finite quark masses. $\Delta/T^4$ goes to zero at low and large temperatures and presents a peak around the transition density.
In the bulk case, our results for $\Delta/T^4$ are in qualitative agreement with lattice QCD results. Moreover, for ${{\cal U}_{F2}}$ we obtain a good quantitative agreement with lattice data up to temperatures around 250 MeV.   
For different Polyakov loop potentials $\cal U$, we find that as the radii decrease the  peak moves towards lower temperatures  and its height increases. At temperatures below that of the peak the results show a stronger dependence on the system's size and on the choice of the Polyakov loop potential.

In the bulk case, our results for the energy density $\epsilon$, the entropy density $s$ and the specific heat $c_V$ are in qualitative agreement with  previous calculations presented in Ref. \cite{BJ} and with lattice QCD results \cite{Bazavov:2014}. 
At high temperatures,  the curves for all system's radii converge to the bulk ones and approach to the Stefan-Boltzmann limit. 
However,  models with the Polyakov loop potentials of Fukushima tend to the  Stefan-Boltzmann limit for quarks only (without gluons) while models with the polynomial and logarithmic Polyakov loop potentials tend to the  Stefan-Boltzmann limit  including quarks and gluons, as already known from previous works \cite{Fukushima:2008wg,BJ,Bhattacharyya2013}.
In general, $\epsilon$, $s$ and $c_V$ don't change significantly  with the change in volume, except for $c_V$ in the transition region and for $\epsilon$ at temperatures below the transition region.

At high temperatures the speed of sound tends to the ultarelativistic limit of an ideal gas, $c_s = 1/\sqrt{3}$ but at lower temperatures, interactions become relevant and $c_s^2$ decreases significantly. Again, for the bulk case we find a qualitative agreement with lattice QCD results. Notwithstanding,  contrary to previous findings \cite{Bhattacharyya2013}, our results show that the speed of sound doesn't depend too much on the system's size. In fact, small variations are observed only in the transition region.

Two very relevant quantities for finite systems are the surface tension and the curvature energy which  have been calculated  for drops with different sizes. 
We find that  $\alpha_{\mathrm{TOT}}$ is largely dominated by the contribution of strange quarks  (in coincidence with previous results for cold quark matter at very high densities \cite{Lugones:2013bo,Lugones:2017ft}), while $\gamma_{\mathrm{TOT}}$ is controlled mainly by the behavior of $u$ and $d$ quarks.
Both, $\alpha_{\mathrm{TOT}}$ and $\gamma_{\mathrm{TOT}}$ change significantly  with $R$ at all temperatures, specially for small drops with radii below 10 fm. There is also a considerable dependence on the Polyakov loop potential. 
For $T \gtrsim 250$ MeV,  $\alpha_{\mathrm{TOT}}$ and $\gamma_{\mathrm{TOT}}$ grow proportionally to $T^{3/2}$. 
At lower temperatures $\alpha_{\mathrm{TOT}}$ has a minimum related to the chiral transition of $s$ quarks and $\gamma_{\mathrm{TOT}}$ has a minimum associated with the $u$ and $d$ quarks chiral transition.
For smaller temperatures,  $\alpha_{\mathrm{TOT}}$ and $\gamma_{\mathrm{TOT}}$ tend to  constant values of the same order of the ones obtained for very dense cold quark matter \cite{Lugones:2013bo}. 

In summary, our main conclusion is that several thermodynamic  quantities are sensitive to finite size effects, particularly  for temperatures around the crossover transition and for systems with radii below $\sim 10$ fm. 
These results can be potentially relevant for the study of the QCD transition at the early Universe  \cite{Roque:2013ufa,MouraoRoque:2017xxo} and should be extended to other regions of the QCD phase diagram, specially the region of high temperatures and moderate baryon chemical potentials where heavy ion collisions take place.

\begin{acknowledgements}
G.L. acknowledges the Brazilian agencies Conselho Nacional de Desenvolvimento Cient\'{\i}fico e Tec\-no\-l\'ogico (CNPq) and Funda\c c\~ao de Amparo \`a Pesquisa do Estado de S\~ao Paulo (FAPESP) for financial support. 
\end{acknowledgements}

\end{document}